\documentclass[preprint2]{aastex}

\newcommand{\hi}{\mbox{H{\sc i}}}
\newcommand{\hii}{\mbox{H{\textsc{ii}}}}
\newcommand{\ha}{H$\alpha$}
\newcommand{\pa}{$P.A.$}

\newcommand{\kms}{km s$^{-1}$}

\newcommand{\msol}{\rm M$_\odot$}

\slugcomment{To appear in AJ}

\shorttitle{KAT--7 Science Verification}
\shortauthors{Carignan et al.}

\begin{document}


\title{KAT-7 Science Verification: Using $\hi$ Observations of NGC 3109 to Understand its Kinematics and Mass Distribution}

\author{C. Carignan$^1$, B. S. Frank, K. M. Hess, D. M. Lucero and T. H. Randriamampandry}
\affil{Department of Astronomy, University of Cape Town, Private Bag X3, Rondebosch 7701, South Africa}

\and

\author{S. Goedhart and S. S. Passmoor}
\affil{SKA South Africa, The Park, Park Road, Pinelands, 7405, South Africa}
\email{ccarignan@ast.uct.ac.za}

\altaffiltext{1}{South African SKA Research Chair in Multi-Wavelength Astronomy}

\begin{abstract}
\hi\ observations of the Magellanic-type spiral NGC 3109, obtained with the seven dish Karoo Array Telescope (KAT--7), are used to analyze its mass distribution. Our results are compared to what is obtained using VLA data. KAT--7 is the precursor
of the SKA pathfinder MeerKAT, which is under construction. The short baselines and low system temperature of the telescope make it sensitive to large scale low surface brightness emission.  The new observations with KAT--7 allow the measurement of the rotation curve of NGC 3109 out  to 32\arcmin , 
doubling the angular extent of existing measurements. A total \hi\ mass of 4.6 $\times 10^8$ \msol\
is derived, 40\% more than what was detected by the VLA observations. 

The observationally motivated pseudo-isothermal {\it dark matter} (DM) halo model can reproduce very well the
observed rotation curve but the cosmologically motivated NFW DM model gives a much poorer fit to the data.
While having a more accurate gas distribution has reduced the discrepancy between the observed RC and the 
MOdified Newtonian Dynamics (MOND) models, this is done at the expense of having to use 
unrealistic mass--to--light ratios for the stellar disk
and/or very large values for the MOND {\it universal} constant $a_0$. Different distances or $\hi$ contents cannot reconcile MOND with the observed kinematics,
in view of the small errors on those two quantities. As for many slowly rotating gas--rich galaxies studied recently,
the present result for NGC 3109 continues to pose a serious challenge to the MOND theory.
\end{abstract}

\keywords{techniques: interferometric -- 
galaxies: individual: NGC 3109 -- galaxies: kinematics and dynamics --
galaxies: haloes -- cosmology: dark matter}

\section{Introduction}
\label{sec:intro}

On 2012 May 25, South Africa was awarded the construction of  the mid-frequency array of the Square Kilometer Array (SKA),
while Australia will build the low-frequency array. The SKA will consist of an
ensemble of 3000 $\sim$15 m dishes of which 80\% will constitute the core portion in the Karoo desert while the
remaining antennae will extend all the way to the 8 African partner countries, namely Botswana, Ghana, Kenya, Madagascar, 
Mauritius, Mozambique, Namibia and Zambia. It is expected that the full SKA will be completed around 2025. 
A precursor array of 64 dishes, MeerKAT, is already under construction by South Africa and should be
ready for science operation 
in 2016. In preparation for these two large projects, a pre-precursor array comprising 7 dishes, KAT--7, was completed
in December 2010. While its main purpose is to test technical solutions for MeerKAT and the SKA, scientific 
targets such as NGC 3109 were also observed during commissioning to test the \hi\ spectral line mode.
In this paper, we compare over 100 hours of observations taken with KAT--7 to previously obtained VLA data
and perform a thorough analysis of the mass distribution of NGC 3109.

NGC 3109 is an SB(s)m galaxy \citep{dev91} on the outskirts of the Local Group \citep{vdb94}. It is even believed 
by certain authors to belong to the Local Group \citep[e.g.][]{mat98}. While NGC 3109 looks like an Irregular galaxy 
on short exposures \citep{san61}, it is clearly a spiral on longer exposures \citep{car85}. Spiral arms are clearly 
visible, especially on the east side. This small spiral (scale length $\alpha^{-1}$ = 1.2 kpc) is a Low Surface Brightness (LSB)
system with $B(0)_c$ = 23.17 \citep{car85}. Its optical parameters are summarized in Table~\ref{optpar}. 
One important parameter for mass modeling is the distance. Fortunately, because of its proximity, numerous Cepheids 
were observed in this Magellanic-type spiral, The most recent measurements are summarized in Table~\ref{dist}. For 
this study, we adopt a distance of 1.30 $\pm\ 0.02$ Mpc \citep{sos06}.

\begin{table}[h!]
\begin{center}
\caption{Optical parameters of NGC 3109 (DDO 236).}
\label{optpar}
\begin{tabular}{lcl}
\tableline\tableline
Parameter  & & Ref \\          
\tableline
Morphological type                                          & SB(s)m                                            & (1)   \\
Right Ascension (J2000)                                &10$^{\rm h}$ 03$^{\rm m}$ 06.7$^{\rm s}$    & (1)   \\
Declination (J2000)                                        & --26$^{\rm o}$  09\arcmin\ 32\arcsec   & (1)   \\
Distance Modulus $(m-M)_0$          & 25.57  $\pm\ 0.02$                & (2)  \\
Distance (Mpc)                                                 & 1.30 $\pm\ 0.02$                            & (2) \\
Scale (pc arcmin$^{-1}$)                              & 378                                                      & \\
Isophotal major diameter, $D_{25}$  & 14.4\arcmin & (3) \\
Holmberg radius, $R_{HO}$  & 13.3\arcmin & (3) \\
Total  apparent $B$ magnitude			               & 10.27  & (4) \\
Corrected apparent $B$ mag.                                    &  9.31  & (1) \\
Absolute $B$ magnitude                                                    &  --16.26           & (2) \\ 
Exponential disk parameters:&&\\
Central $B(0)_c$               & 23.17 & (4) \\
Scale length, $\alpha^{-1}$ (kpc)                                        &  1.2       & (4) \\
\hline
\tablerefs{
(1) \citet{dev91}; \\ (2) \citet{sos06}; (3) \citet{jc90}; \\ (4) \citet{car85}.}
\end{tabular}
\end{center}
\end{table}

\begin{table}[h!]
\begin{center}
\caption{Cepheids distance estimates for NGC 3109.}
\label{dist}
\begin{tabular}{lcc}
\hline\hline
Reference  & (Mpc)  \\          
\hline
\citet{sos06}        & 1.30 $\pm\ 0.02$  \\ 
\citet{pie06}        & 1.28 $\pm\ 0.03$  \\ 
\citet{mus97}      & 1.36 $\pm\ 0.10$  \\ 
\citet{cap92}  & 1.26 $\pm\ 0.10$  \\    
Mean Cepheids distance & $< 1.30\ \pm 0.04>$ \\                                          
\hline
\end{tabular}
\end{center}
\end{table}

NGC 3109  is of significant scientific interest for two main reasons. Firstly, \citet{jc90} used observations 
with the hybrid VLA DnC configuration (synthesized beam of 36\arcsec $\times$  27\arcsec\   and velocity resolution of 10.3 \kms) to perform a dynamical 
study of this galaxy, comparing the rotation curve (RC) derived from a tilted-ring analysis (see Sec.~\ref{sec:tr})  to models composed of a 
luminous disk (stars \& gas) and of a dark isothermal (ISO) halo (see Sec.~\ref{sec:iso}). Such a mass model provides an excellent fit
to this nearly solid-body type RC. Combining the \hi\ RC with higher resolution \ha\ kinematical data, \citet{sbo01}
also obtained a very good ISO Dark Matter (DM) model but a much less accurate fit for the cosmologically motivated
NFW \citep{nfw97} DM model. 

Recently, it was also shown that a MOdified Newtonian Dynamics \citep[]{mil83,mil88} model  (MOND) 
could  not reproduce  the NGC 3109's RC \citep{ran13}, at least with the data available. With our new data, it should be possible 
to compare DM (ISO and NFW) models to MOND (no dark matter) models and see if NGC 3109
really challenges the MOND theory. This is not the first time that NGC 3109 poses problems to MOND \citep[see e.g.][]{san86,bbs91}.

Secondly, \citet{bdb01} used 21cm Multibeam data with the Parkes 64m dish (beam $\sim$ 15.5\arcmin\  and velocity
resolution of 1.1 \kms) to study the environment of NGC 3109. They provide a compelling argument that the warp  in the \hi\  disk of 
NGC 3109 could be due to a dynamical encounter with the Antlia dwarf. This is also suggested by the elongation
of the optical isophotes of NGC 3109 toward the south \citep{jc90} and those of Antlia in the direction of NGC 3109 \citep{pen12}. 

However, recent derivation of various merger and/or interaction parameters (e.g. asymmetry, clumpiness) by \citet{pc12}  are consistent
with Antlia being an undisturbed dwarf elliptical. In fact, despite its dSph appearance, Antlia is better classified as
a dSph/dIrr transition type \citep{ggh03} because of its high \hi\ content. With high sensitivity and mainly
better spatial resolution observations (KAT--7 vs HIPASS), it should be possible to map much better the traces of that interaction, if it exists.
An encounter/interaction that has significantly altered NGC3109's kinematics would give less weight to the finding that MOND 
cannot reproduce the rotation curve.

The remainder of this paper is as follows. 
In Sec.~\ref{sec:new}, a description of the new radio interferometer KAT-7 is given. Sec.~\ref{sec:KAT7} describes in details 
the new \hi\ data obtained with KAT--7,  Sec.~\ref{sec:ANGST} those from the VLA--ANGST survey and Sec.~\ref{sec:comp}
compares the different data sets. Sec.~\ref{sec:rc} derives the optimal RC that is used for the DM
(ISO and NFW) and MOND models of Sec.~\ref{sec:mass}. A discussion follows
in Sec.~\ref{sec:discussion} and a summary of the results and the final conclusions are given in Sec.~\ref{sec:conclusion}.

\section{A New Radio Interferometer: KAT--7}
\label{sec:new}

\begin{figure}[h!]
\includegraphics[width=\columnwidth]{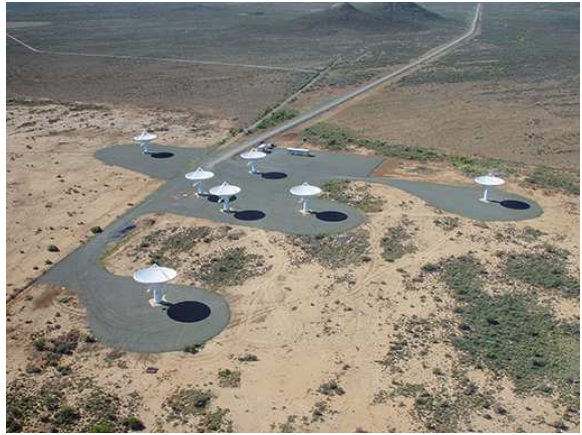}
\caption{Aerial view of the KAT--7 array in the Northern Cape's Karoo desert, South Africa.}
\label{fig:kat7}
\end{figure}

The seven-dish KAT--7 array, shown in Fig.~\ref{fig:kat7}, was built as an engineering testbed for the
64-dish Karoo Array Telescope, known as MeerKAT, which is the
South African pathfinder for the Square Kilometer Array (SKA).  KAT--7
and MeerKAT are located close to the South African SKA core site in
the Northern Cape's Karoo desert region.  KAT--7 is remotely controlled
from Cape Town, some 800 km away from the site. Construction of the array started in early 2008 and was 
completed in December 2010, with "first light" fringes obtained between two antennas in December 2009.
The instrument is now in its science verification stage. 

The array is extremely compact, with baselines ranging between 26 m to 185 m.
The KAT-7 layout was determined using the optimization algorithm described in \citet{dev07}, which determined a layout with a Gaussian UV distribution for a specified observation setting. The observation setting being optimized in this case was an 8 hour track (symmetric hour angle range), on a target at a --60 degree declination. The optimization objective was a Gaussian UV distribution at 1.4GHz, yielding a Gaussian synthesized beam with low sidelobes.  Several randomly seeded layouts were generated and were evaluated for a set of observation options (time durations: snapshot, 4hr, 8hr, 12hr; declinations: 0, --30,--60, --90 degrees). The layout selected had the lowest sidelobes for the largest number of test observation settings considered. The antenna layout can be found at https://sites.google.com/a/ska.ac.za/public/kat-7.

The KAT-7 dishes have a prime-focus alt-az design with a F/D of 0.38,
optimized for single-pixel L-band feeds. The low noise amplifiers
(LNAs) for  the feeds are cryogenically cooled to 80 K using Stirling
coolers. The key system specifications for KAT--7 are summarized in
Table~\ref{K7spec}. The digital backend of the system is an FPGA 
(Field Programmed Gate Array)-based, flexible packetised correlator
using the
Reconfigurable Open Architecture Computing Hardware 
 (ROACH: https://casper.berkeley.edu/wiki/ROACH), which is a
flexible and scalable system enabling spectral line modes covering a
wide range of resolutions. Table~\ref{K7corr} gives the details of
the recently commissioned correlator modes. Digital filters give a
flat bandpass over the inner 75\% of the band with a rapid roll-off at
the edges of the band.

\begin{table}[h!]
\begin{center}
\caption{KAT--7 specifications.}
\label{K7spec}
\begin{tabular}{lc}
\hline\hline
Parameter & Value \\
\hline
Number of antennas 	& 7\\
Dish diameter			& 12 m \\
Min baseline				& 26 m\\
Max baseline				& 185 m\\
Frequency range		& 1200 - 1950 MHz \\
Max instantaneous bandwidth & 256 MHz \\
Polarisation & Linear H \& V \\
T$_{\rm sys}$							& 26 K \\
Aperture efficiency   & 0.65 \\
System Equivalent Flux Density & 1000 Jy \\
Latitude						&  -30:43:17.34 \\
Longitude					& 21:24:38.46 \\
Elevation					& 1038 m \\
Digital back-end			& ROACH boards\\
\hline
\end{tabular}
\end{center}
\end{table}

\begin{table}[h!]
\begin{center}
\caption{KAT--7 correlator modes.}
\label{K7corr}
\begin{tabular}{lccc}
\hline\hline
mode & total BW & number of  & channel BW \\
		&	(MHz)				& 		channels				& (kHz)\\
\hline
c16n2M4k & 1.5625		& 4096							& 0.381 \\
c16n7M4k & 6.25			& 4096							& 1.526 \\
c16n25M4k & 25			& 4096							& 6.104 \\
c16n400M4k & 256		& 1024 					& 390.625 \\
\hline
\end{tabular}
\end{center}
\end{table}

CASA ({\it Common Astronomy Software Applications}; \citealt{mcm07}) 
is the standard data reduction package being used for the reduction
of the KAT--7 data and is anticipated to be used for MeerKAT.

\section{$\hi$ Observations of NGC 3109}
\label{sec:HI}

The KAT--7 \hi\  observations of NGC 3109 provide a unique opportunity to simultaneously achieve \hi\ spectral-line 
science verification and an original scientific result. They  complement the high
spatial resolution ($\sim$ 10\arcsec) but small field of view ($\sim 30\arcmin$) of the VLA--ANGST data 
\citep{ott12} and the high sensitivity ($\sim 10^{17}\ \rm{cm^{-2}}$) but low spatial resolution ($\sim 15.5\arcmin$) 
Multibeam data \citep{bdb01}.
With its short baselines and low system temperature (T$_{\rm sys} \sim 26$K), KAT--7
is very sensitive to low surface brightness and large scale \hi\ emission,
characteristic of the signal expected from NGC 3109.

\subsection{KAT--7 data on NGC 3109}
\label{sec:KAT7}

In order to observe the \hi\ in both NGC 3109 and Antlia, plus possible signs of interaction between the two,
a mosaic of 3 fields was obtained to have good sensitivity over a region of $1.5^{\rm o}{\rm (EW)} \times 3^{\rm o}{\rm (NS)}$.
The data was collected over 13 observing sessions between 2012 November 20 and 2012 December 26 using the 
c16n7M4k spectral line mode (Table~\ref{K7corr}) for a median of 11 hours in each session and a total 
of 122$^{\rm h}$43$^{\rm m}$56$^{\rm s}$, including calibration. This yielded a total time on source of $\sim25$ hours for each pointing. 

The first three sessions were taken with 6 cold antennae, but the entire array was available for the remaining 10 observing 
sessions.  The roughly 1 degree beam of KAT-7 is just large enough to image NGC 3109 in a single pointing. We used 
three pointings positioned in a straight line and extending slightly to the SE to mosaic the region 
between NGC 3109 and Antlia.  The distance between pointings was chosen to give a uniform coverage between the phase centers.  
The c16n7M4k correlator mode gives velocity
channels of 0.32 \kms\ over a flat bandpass of $\sim1000$ \kms\ , centered at 1417 MHz. The large bandwidth allows 
to collect \hi\ data on background galaxies in the field. 

The basic data reduction was done in CASA 3.4.0 and 4.0.0. More advanced analysis
was done using either AIPS \citep{gre03}, MIRIAD \citep{stw95} and/or GIPSY \citep{vdh92}.  
To start with, the data was flagged in an automated way to discard data for shadowing and 
flux calibrators below 20 degrees in elevation.  The data was additionally examined as a function 
of frequency and baseline, and flagged by hand.  

This testing of the HI spectral line mode led to 
the discovery of faint, very narrow, internally generated radio frequency interference (RFI) originating 
along the signal path, which has since been successfully eliminated in KAT-7 by the insertion of 
a low-pass filter.  The RFI in our data is antenna dependent and only affects about 30 channels out of the central 3000 on three antennae.
One of the primary goals of the {\it science verification phase} is exactly to identify these type of problems and correct for them.

The standard interferometric data reduction strategy that has been employed for decades in AIPS and Miriad has been used.  
Each of the 13 observing sessions was reduced individually.  Continuum subtraction was accomplished by selecting line free channels 
and using a first order fit.  KAT--7 does not use Doppler tracking, and CASA does not fully recognize frequency keywords, 
so special care was taken to produce image cubes with the proper velocity coordinates.  This was a three steps process accomplished by:
\begin{itemize}
\item setting the MEAS\_FREQ\_REF and REF\_\\FREQUENCY keywords in the SPECTRAL WINDOW table, 
\item specifying the reference frequency and setting the output frame to optical, barycentric in CVEL, and 
\item specifying the rest frequency again in the task CLEAN.
\end{itemize}

The calibration was applied and the three mosaic pointings were then SPLIT from the calibration sources. The data were averaged in time 
from 5 to 10 second integrations, and spectrally from 0.32 \kms\ to 1.28 \kms\ channels.  All 13 data sets were then combined in CONCAT.

The data was imaged using the mosaic mode and the multi--scale clean option.  Three cubes were produced  by applying  
natural (na), uniform (un), 
and neutral (ro: robust=0) weighting (Table~\ref{kat7par}) to the uv data. 
The robust=0 cube was cleaned interactively using a mask to select regions of galaxy emission by hand 
in each channel.  After each major clean cycle, the mask was expanded to include regions of fainter galaxy 
emission.  After a final mask was created, the cube was reproduced using the final mask in a non-interactive 
clean down to the noise threshold.  All cubes and images were produced using the same mask derived from 
the robust=0 weighted cube.  This provided a compromise between surface brightness sensitivity to large-scale 
emission and a low noise threshold, while mitigating confusion from sidelobes and low-level artifacts due to instrument calibration.

\begin{table}[h!]
\begin{center}
\caption{Parameters of the KAT--7  observations.}
\label{kat7par}
\begin{tabular}{lr}
\hline\hline
Parameter & Value \\
\hline
Start of observations & 20 nov. 2012 \\
End of observations & 26 dec. 2012 \\
Total integration per pointing &  24.75 hours\\
FWHM of primary beam & 58.67\arcmin \\
Total Bandwidth & 6.25 MHz \\
Central frequency &  1417 MHz \\
Channel Bandwidth (4 x 1.526 kHz) & 6.1 kHz \\
Number of channels (4096/4) & 1024 \\
Channel width (4 x 0.32 \kms) & 1.28 \kms\\
Maps gridding & 55\arcsec\ x 55\arcsec \\
Maps size & 128 x 256 \\
Flux/bandpass calibrator & 3C138 \\
Phase calibrator & 0118-317 \\
\hline
Robust = 0 weighting function & \\
FWHM of synthesized beam & 212\arcsec  x 201\arcsec \\
RMS noise  & 5.6 mJy/beam \\
Conversion $^{\rm o}$K/(1 mJy/beam) & 0.014 \\
\hline
Natural weighting function & \\
FWHM of synthesized beam &  264\arcsec  x 233\arcsec \\
RMS noise (mJy/beam)  & 3.7 mJy/beam \\
Conversion $^{\rm o}$K/(1 mJy/beam) & 0.010 \\
\hline
Uniform weighting function & \\
FWHM of synthesized beam & 203\arcsec  x 196\arcsec \\
RMS noise  &  9.1 mJy/beam \\
Conversion $^{\rm o}$K/(1 mJy/beam) & 0.015  \\
\hline
\end{tabular}
\end{center}
\end{table}

\begin{figure}[h!]
\includegraphics[width=\columnwidth]{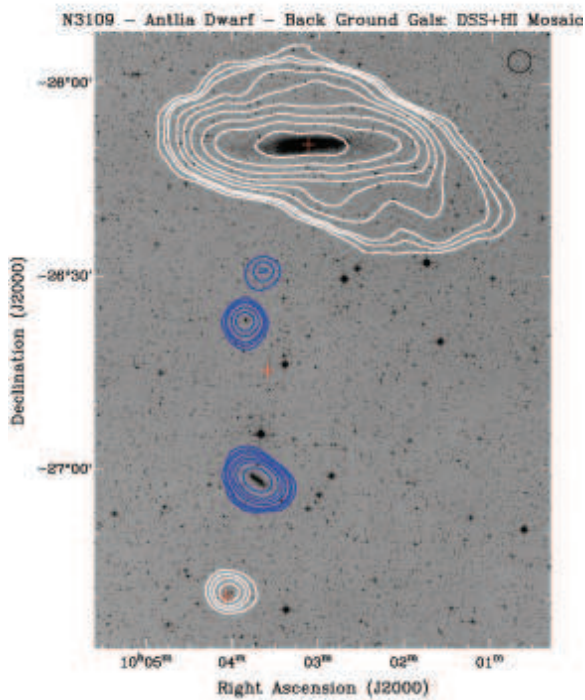}
\caption{Sum of the channel maps of the KAT-7 \hi\ mosaic. The centers of the 3 fields are shown with red crosses.
The contours are 0.1, 0.2, 0.3, 0.6, 1.0, 1.6, 3.2, 6.4, 12.8 $\times 10^{20}$ atoms cm$^{-2}$, superposed on a DSS B image. 
NGC 3109 ($V_{sys}$ = 404 \kms) at the top and Antlia ($V_{sys}$ = 360 \kms) at the bottom are shown with white contours. 
The two background galaxies ESO 499--G037
($V_{sys}$ = 953 \kms), ESO 499--G038 ($V_{sys}$ = 871 \kms) and its associated \hi\ cloud ($V_{sys}$ = 912 \kms)
are shown with blue contours. The synthesized beam is shown in the upper-right corner.}
\label{fig:kat7_squash_B}
\end{figure}

In addition to NGC 3109, Antlia, ESO 499-G037 and ESO 499-G038, an HI cloud which has no known optical counterpart
was serendipitously discovered to the north of 
ESO 499-G038 at a similar velocity (Figure~\ref{fig:kat7_squash_B}).  The channels 
that contain this emission 
are remarkably clean and uniform in their noise characteristics.  By contrast the channels which contain the brightest HI emission 
from NGC 3109 contain artifacts from sidelobes of the telescope's synthesized beam, which we have been unable to remove completely.  
The noise value in these channels is three times higher than elsewhere in the cube. For our analysis of NGC 3109, the cube
produced with na weighting is used, except for the map showing the sum of all the channels (Figure~\ref{fig:kat7_squash_B})
which used the ro weighting scheme. The parameters of the KAT--7 observations are summarized in Table~\ref{kat7par}.

Fig.~\ref{fig:kat7_squash_B} shows the total intensity map for all the channels of the data cube. The lowest contour is at 1.0 x 10$^{19}$ atoms cm$^{-2}$.
Besides NGC 3109 to the north and Antlia to the south, the two background galaxies 
ESO 499-G037 (10:03:42 -27:01:40; V$_{sys}$ = 953 \kms)
and ESO 499-G038 (10:03:50 -26:36:46; V$_{sys}$ = 871 \kms) are clearly visible. 
More details about these two systems will be given in the Appendix.
As mentioned earlier, the small cloud between ESO 499-G038 and NGC 3109 has no obvious optical counterpart. 
However, it is clearly
associated with ESO 499-G038 and not NGC 3109, being at a velocity $>$ 900 \kms.

The natural weighted cube is the best place to look for low-surface brightness emission between NGC 3109 and Antlia, 
but there is no obvious evidence of it there.  In fact the elevated noise and strong sidelobes in the channels, which 
contain bright NGC 3109 data, prevent us from detecting lower surface brightness emission and limit what we can 
learn from smoothing the data.  This work on NGC 3109 showed that better models of the KAT-7 primary beam are 
needed for calibration before we can go deeper. 

\begin{figure}[h!]
\includegraphics[width=\columnwidth]{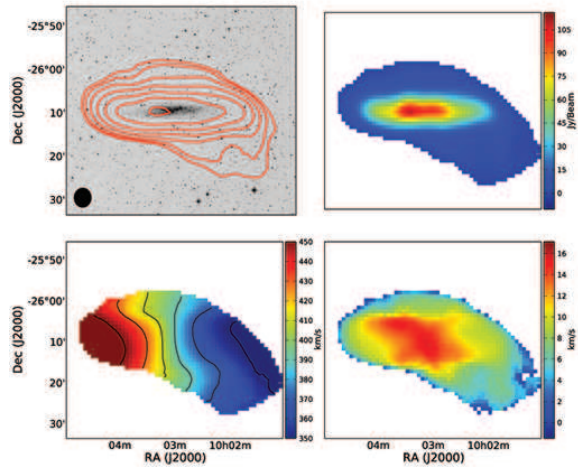}
\caption{Moment maps of NGC 3109 from the KAT-7 na data cube. (top-left): total \hi\ emission map with the synthesize 
beam in the bottom--left corner. 
The contours are 0.4, 0.8, 1.6, 3.2, 6.4, 12.8, 25.6 $\times 10^{20}$ atoms cm$^{-2}$, superposed on a DSS B image. 
(top-right): total \hi\ emission map; (bottom-left): velocity field. The contours are 350, 370, 390, 410, 430, 450 \kms\ ;
(bottom-right): velocity dispersion.}
\label{fig:moments_n3109}
\end{figure}

Fig.~\ref{fig:moments_n3109} shows the result of the moment analysis of the NGC 3109 data. It can be seen that the
\hi\ extends over nearly 1$^{\rm o}$, more than 4 times the optical diameter ($D_{25}$). The nearly parallel isovelocity contours,
typical of a solid--body type rotation curve, are clearly visible, as well as the warp of the \hi\ disk in the outer parts. The
velocity dispersion map shows very well the gradient of $\sigma$ from 15 \kms\ in the center down to 5 \kms\ at the edge of the disk.

\begin{figure}[h!]
\includegraphics[width=\columnwidth]{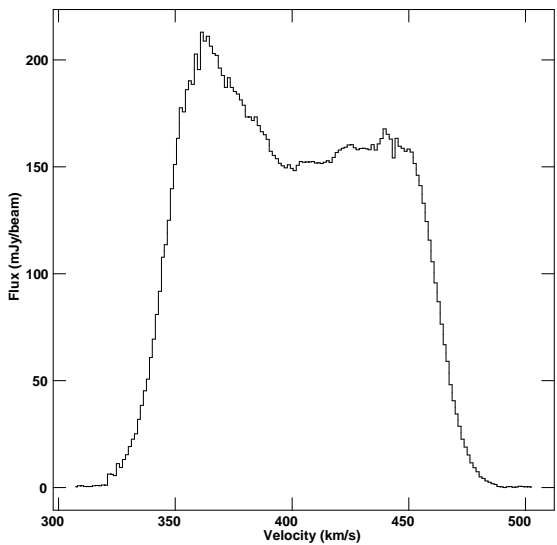}
\caption{Global \hi\ line profile of NGC 3109 from the KAT-7 na data cube.
The small dip $\sim$450 \kms\ is due to internally generated narrow band RFI.}
\label{fig:global_n3109}
\end{figure}

Fig.~\ref{fig:global_n3109} gives the integrated \hi\ profile for NGC 3109.  Profile widths of $\Delta$V$_{50}$ = 
$118 \pm 3$ \kms\ and $\Delta$V$_{20}$ = $136 \pm 3$ \kms\ are derived. Since the \hi\ distribution is clearly lopsided, with more gas
on the approaching (SW) than on the receding side, we adopt 
the midpoint velocity of 404 $\pm$  2 \kms\ as more representative of the systemic velocity than the intensity--weighted mean velocity.
This can be compared to 404 \kms , $\Delta$V$_{50}$ = 123 \kms\ and $\Delta$V$_{20}$ = 137  \kms\ in \citet{jc90}.
An integrated flux of $1142 \pm 110$ Jy \kms\ is measured which corresponds at our adopted distance
of 1.3 Mpc to a \hi\ mass of M$_{\hi} = 4.6 \times 10^8$ \msol\  for a
\hi\ mass--to--luminosity ratio M$_ {\small \hi}$/L$_B$ of 1.0, showing the gas--rich nature of NGC 3109.
With a HPBW of the primary beam of nearly one degree and the short baselines available, no flux should be missed by these observations.

Two methods of kinematical analysis will be used for the NGC 3109 data, namely an intensity-weighted moment analysis
and a Gauss--Hermite polynomial profile fit. With such high S/N data and low velocity gradient RC, not much difference
is expected between the two types of analysis. As a result, the analysis yielding the smaller errors and the largest radius RC will be used for the mass model analysis
in Sec.~\ref{sec:mass}. The RCs are derived in Sec.~\ref{sec:rc}.

\subsection{VLA--ANGST data on NGC 3109 and Antlia}
\label{sec:ANGST}

NGC 3109 and Antlia were observed at the VLA as part of the VLA--ANGST survey \citep{ott12}.
The galaxies were observed for $\sim$9 hours in BnA, $\sim$3 hours in CnB and $\sim$3 hours in DnC
configurations giving access to scales from $\sim$6" to $\sim$15'. The hybrid configurations were used to
get more circular beams for these two southern objects. The data were gridded with two
different weighting functions: na weighting for maximum sensitivity and ro weighting
for maximum spatial resolution (smaller synthesized beam). We use the 
ro weighted maps for NGC 3109 and the na weighted maps for Antlia.
The parameters of the VLA observations are given in Table~\ref{angstpar}.

\begin{table}[h!]
\centering
\caption{Parameters of the VLA--ANGST  observations.}
\label{angstpar}
\begin{tabular}{lcc}
\hline\hline
Parameter   & NGC 3109 & Antlia  \\          
\hline
FWHM of primary beam & 31.5\arcmin & 31.5\arcmin \\
FWHM of synthesized beam & 7.6\arcsec x 5.0\arcsec  & 14.1\arcsec x 13.9\arcsec \\
Total Bandwidth (MHz) & 1.56  & 0.78 \\
Number of channels & 256 & 128 \\
Central frequency (MHz) & 1418.5 & 1418.6\\
Channel width (\kms) & 1.3 & 1.3 \\
RMS noise (mJy/beam)  &  1.7 & 1.0 \\
Conversion $^{\rm o}$K/(1 mJy/beam) & 15.8  &  3.06 \\
Maps gridding & 1.0\arcsec x 1.0\arcsec & 1.5\arcsec x 1.5\arcsec \\
Maps size & 2048$^2$ & 1024$^2$ \\
Weighting function & robust & natural \\
Phase calibrator & 0921-263 & 0921-263\\
\hline
\end{tabular}
\end{table}

\begin{figure}[h!]
\includegraphics[width=\columnwidth]{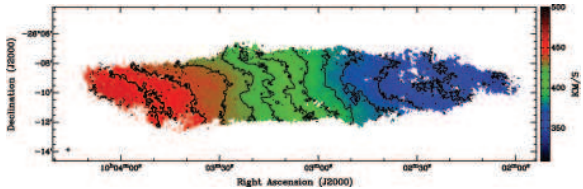}
\caption{Velocity field for the ro-weighted VLA--ANGST velocity field.}
\label{fig:angst_VF}
\end{figure}

Figure~\ref{fig:angst_VF} shows the velocity field obtained with the ANGST data for NGC 3109.
The nearly parallel contours are again clearly visible. 
From the \hi\ emission map, we measure a total flux of 723 Jy \kms\ which, at our adopted distance, corresponds to 
a M$_{\hi} = 2.9 \times 10^8$ \msol, which is nearly 40\% less than the \hi\ detected with KAT--7. 
Because the VLA is not sensitive to scales larger than 15\arcmin ,
the VLA-ANGST data will be missing some flux and their \hi\ mass will be clearly underestimated
for NGC 3109. In fact, the \hi\ mass measured is exactly the same as that found by \citet{jc90} with a VLA
mosaic of two fields. So the problem with the VLA data is not as much the smaller HPBW of the antennae but more
the lack of short baselines.

\begin{figure}[h!]
\includegraphics[width=\columnwidth]{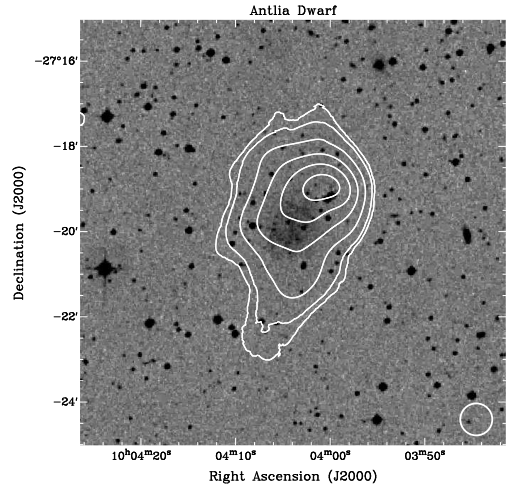}
\caption{Total VLA--ANGST \hi\ emission map of Antlia. The na-weighted data have been spatially smoothed 
to 45\arcsec\ (see the beam in the bottom right corner).
The contours are at $0.3, 0.45, 0.9, 1.5, 2.1\  \&\  2.7  \times 10^{20}$ atoms cm$^{-2}$.
The contours are superposed on a DSS B image.}
\label{fig:angst_antliam0B}
\end{figure}

The VLA--ANGST data, with its high spatial resolution, is much better suited to study the \hi\ distribution in Antlia,
since KAT--7 has barely two beamwidths across the object.
Figure~\ref{fig:angst_antliam0B} gives the total \hi\ emission map for Antlia, where the lowest  contour corresponds to 
$3.0 \times 10^{19}$ atoms cm$^{-2}$. A total flux of  $3.7 \pm 0.4$ Jy \kms\ is found which, at an adopted distance of $1.31 \pm 0.03$ Mpc
\citep{pc12}, corresponds to a M$_{\hi} = 1.50 \pm 0.15 \times 10^6$ \msol. This time, most of the flux should have been detected since
Antlia is much smaller than NGC 3109. However, this total flux is twice as much as that of the HIPASS data where \citet{bdb01} found
$1.7 \pm 0.1$ Jy \kms\ and still 40\% more than the value of $2.7 \pm 0.5$ Jy \kms\ found by \citet{fou90}. The reason for this large difference
is not clear. But since we do not have access to the raw data, it is difficult for us to investigate further.

\begin{figure}[h!]
\includegraphics[width=\columnwidth]{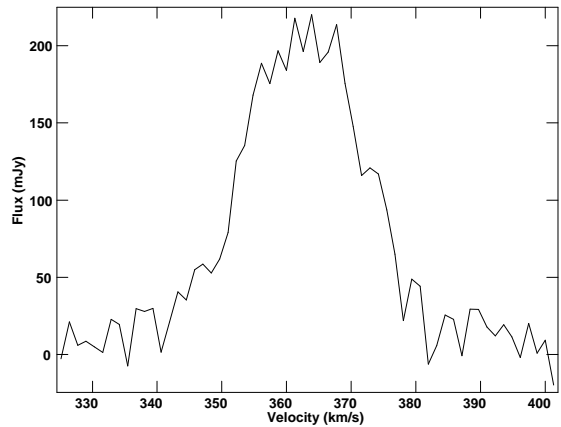}
\caption{VLA--ANGST global \hi\ profile of Antlia using the na-weighted data smoothed 
to 45\arcsec.}
\label{fig:angst_antlia_gp}
\end{figure}

Figure~\ref{fig:angst_antlia_gp} gives the integrated \hi\ profile for Antlia. An integrated flux of $3.7 \pm 0.30$ Jy \kms\ is found, which
is similar to the flux derived from the \hi\ emission map. An intensity-weighted mean velocity of $360 \pm 2$ \kms\ is 
derived along with $\Delta$V$_{50}$ = 
$23 \pm 3$ \kms\ and $\Delta$V$_{20}$ = $33 \pm 3$ \kms. This can be compared to $362 \pm 2$ \kms\ and $\Delta$V$_{20}$ = $30 \pm 2$ \kms\ for \citet{bdb01}
and $361 \pm 2$ \kms\,  $\Delta$V$_{50}$ = $21 \pm 4$ \kms\ and a $\Delta$V$_{20}$ = $33 \pm 5$ \kms\ for \citet{fou90}. 

\subsection{Comparison of the different \hi\ data sets}
\label{sec:comp}

The limiting surface densities of the different interferometric studies are given in Table~\ref{HIsd}.
As far as the VLA data are concerned, one should not be surprised that the \citet{jc90} data 
go deeper than the ANGST data since they are a mosaic
of 2 fields with the same observing time in DnC configuration than the single field ANGST data. 
The 1.3 \kms\ resolution ANGST data is useful for comparisons with the  \citet{jc90} 
10.3 \kms\ resolution data in the inner regions of NGC 3109.  
But mainly, that data provide more information on the \hi\ distribution of Antlia, for which the KAT--7 data is of too low
spatial resolution.

\begin{table}[h!]
\centering
\caption{Limiting \hi\ surface densities of the different interferometer studies for NGC 3109.}
\label{HIsd}
\begin{tabular}{lcc}
\hline\hline
Reference   & atoms cm$^{-2}$ & \hi\ size\\          
\hline
KAT-7         & $1.0 \times 10^{19}$ & 58\arcmin\ x 27\arcmin \\  
ANGST, VLA    & $2.0 \times 10^{20}$  & 32\arcmin\ x 08\arcmin \\ 
\citet{bdb01} &   $2.0 \times 10^{17}$ & 85\arcmin\ x 55\arcmin \\ 
\citet{jc90} & $1.0 \times 10^{19}$  & 40\arcmin\ x 12\arcmin \\                            
\hline
\end{tabular}
\end{table}

In view of the surface densities limits, we see that while both the KAT-7 and the \citet{jc90} reach $1.0 \times 10^{19}$ cm$^{-2}$,
the KAT--7 data covers a larger area since it is sensitive to large scales invisible to the VLA. As for the \citet{bdb01} data
reaching the much lower surface densities of $2.0 \times 10^{17}$ cm$^{-2}$, they provide
the largest detected size (85\arcmin x 55\arcmin). However, this increase in size may be due partly to the large $\sim$15.5\arcmin\  HIPASS beam.

The \hi\ mass estimates of both the single dish and the interferometric observations can be found in Table~\ref{HImass}.
The first thing to notice is the larger mass obtained by the single dish observations compared to the aperture
synthesis ones, the only exception being the \citet{wg77} data which come from a single pointing of the
Parkes 64 m radio telescope. With a $\sim$15\arcmin\ beam, necessarily a lot of the flux extending over 
$\sim$ 1$^{\rm o}$ has been missed. For the others, the discrepancies can be explained by either the way the multi-pointing
data have been combined or most likely that the correction for self--absorption that most of these authors
have applied has been overestimated. No such correction has been applied to the synthesis data.

\begin{table}[h!]
\centering
\caption{Different \hi\ mass estimates for NGC 3109\tablenotemark{*}.}
\label{HImass}
\begin{tabular}{ll}
\hline\hline
Reference   & $10^8$  \msol  \\          
\hline
Aperture synthesis & \\
\hline
KAT-7         & 4.6 $\pm\ 0.5$  \\  
ANGST, VLA           & 2.9 $\pm\ 0.3$  \\ 
\citet{bdb01} & 4.5 $\pm\ 0.6$  \\ 
\citet{jc90} & 2.9 $\pm\ 0.6$  \\  
\hline
Single--dish & \\
\hline
Huchtmeier et al. (1980) & 6.6 $\pm\ 0.7$  \\ 
\citet{wg77}   & 2.1  $\pm\ 0.2$  \\
\citet{dd75} & 5.9 $\pm\ 0.3$ \\ 
\citet{huc73} & 5.9  \\ 
\citet{vd66} & 5.9  \\ 
\citet{eps64}   & 6.6  $\pm\ 0.8$  \\                             
\hline
\tablenotetext{*}{Naturally, all the masses have been corrected to our adopted \\
distance of 1.3 Mpc.}
\end{tabular}
\end{table}

As for the synthesis data, we see that both sets of VLA data agree exactly. We would have expected some more flux
from the deeper  \citet{jc90} data but since most of the flux is in the bright central components, 
the difference is probably just of the order of the errors. On the other hand, both the KAT-7 data and the HIPASS
data agree very well which is surely indicative that, in both cases, no flux is missed. Because both those
data sets see all the scales, they detect nearly 40\% more flux than the VLA, which do not have the proper
short spacings to see scales larger than 15\arcmin .

\subsection{Derivation of the optimal RC}
\label{sec:rc}

The same method is used to derive the RC for both the VLA--ANGST and the KAT--7 data sets. 
For this study, we used the implementation of the tilted ring model in the GIPSY task ROTCUR \citep{beg89}.

\subsubsection{Tilted Ring Models}
\label{sec:tr}

A set of concentric rings is used to describe the motion of the gas in the galaxy. The gas is assumed to be in circular motion. Each ring is characterized by 
a set of 5 orientation parameters, namely: a rotation centre $(x_c,y_c)$, a systemic velocity $V_{sys}$, an inclination $i$, a Position Angle $PA$ and by
a rotation velocity $V_{C}$. Naturally, the rotation centre $(x_c,y_c)$ and the systemic velocity $V_{sys}$ should be the same for all the rings
but $i$ and $PA$ will vary if the \hi\ disk is warped.

The line of sight velocity at any $(x,y)$ position in a ring with radius $R$ is given by

\begin{equation}
\label{ }
V(x,y) = V_{sys} + V_{C} sin(i) cos(\theta) 
\end{equation}
 where $\theta$ is the position angle with respect to the receding major axis measured in the plane of the galaxy. $\theta$ is related to the 
 actual $PA$ in the plane of the sky by
 \begin{equation}
\label{ }
cos(\theta) = \frac{-(x  - x_{0}) sin(PA) + (y - y_{0}) cos(PA)}{R} \\\\
\end{equation}
\begin{equation}
\label{ }
sin(\theta) = \frac{-(x  - x_{0}) cos(PA) + (y - y_{0}) cos(PA)}{R cos(i)}
\end{equation}

A $ |cos{\theta}|$ weighting function and an exclusion angle of $\pm 15\deg$ about the minor axis have been used to give maximum weight 
to the velocity points close to the major axis and minimize the influence of large deprojection errors close to the minor axis 
in view of the large inclination of the galaxy. The width of the rings has been matched to the
synthesized beam size to make sure that the velocity points are independent.  

The method consists at finding for each ring the best set of the 5 orientation
parameters $(x_c,y_c)$,  $V_{sys}$, $i$ and $PA$ 
which minimizes the dispersion of $V_C$ inside the ring.
The following procedure is followed:
\begin{itemize}
\item The rotation center $(x_c,y_c)$ and the systemic velocity $V_{sys}$ are determined first
by keeping $i$ and $PA$ fixed (usually using the optical values). The rotation center and
the systemic velocity have to be determined simultaneously since they are correlated. They
are searched using only the central rings (e.g. $R \le R_{25}$) since there is usually no warp within the optical disk.
\item Now, keeping $(x_c,y_c)$ and $V_{sys}$ fixed, $i$ and $PA$ are looked for to map the
warp of the \hi\  disk, which
usually starts just outside the optical. Here also, $i$ and $PA$ have to be determined
simultaneously since they are also correlated.
\item The previous two steps were done using the data of all the galaxy. Using the same fixed
$(x_c,y_c)$ and $V_{sys}$, the previous step is repeated for the approaching and
receding sides separately to see possible departures from axisymmetry.
\end{itemize}

The errors on $V_C$ will be the quadratic sum of the dispersion $\sigma$ in each ring
and half the difference between the approaching and the receding sides:

\begin{equation}
\Delta V = \sqrt {\sigma ^2(V) + ({{ |V_{app} - V_{rec} |}\over 2})^2} 
\end{equation}
 
Since the mass models assume an axisymmetric system, we think that this way of
calculating the errors is more representative of the true uncertainties, when comparing the RC
to the model.

\subsubsection{VLA--ANGST RC}

\begin{figure}[h!]
\includegraphics[width=\columnwidth]{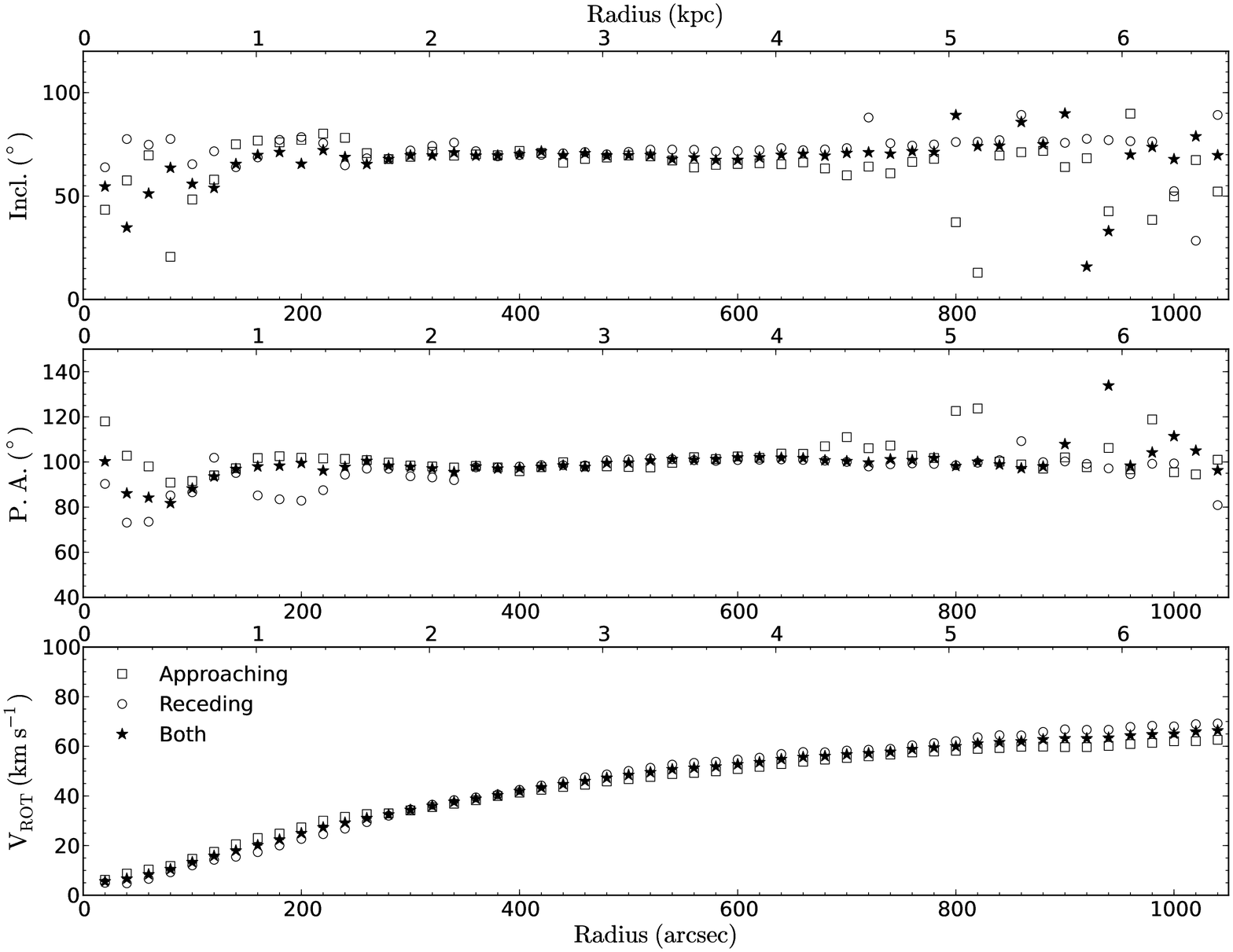}
\caption{Tilted-ring model using the ro VLA--ANGST data for both sides and independently 
for the approaching and the receding side.}
\label{fig:tr_angst}
\end{figure}

\begin{figure}[h!]
\includegraphics[width=\columnwidth]{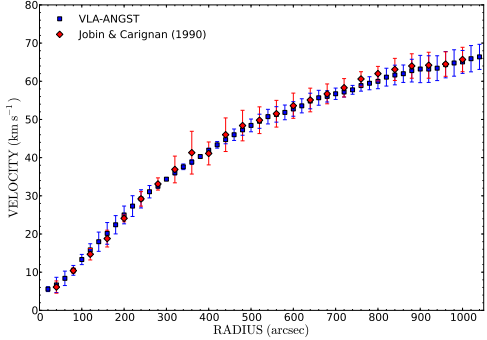}
\caption{Comparison of the VLA--ANGST and the  \citet{jc90} rotation curves.}
\label{fig:tr_angst_jc90}
\end{figure}

For the derivation of the kinematics of NGC 3109, we use the ro-weighted VLA--ANGST
data. We find that the rotation center is coincident with the optical center and derive a 
systemic velocity $V_{sys} = 402$ \kms. By keeping those parameters fixed, we then fitted $i$ and $PA$. 
The solutions for the whole galaxy and separately for the approaching and receding sides are shown in Fig.~\ref{fig:tr_angst},
along with a comparison with \citet{jc90} in Fig.~\ref{fig:tr_angst_jc90} . We see that the warp
which starts around the Holmberg radius ($R_{HO}$), is more 
important on the approaching (SW) side and that the RC from both VLA datasets agree very well,
despite the difference in velocity resolutions (1.3 \kms\ vs 10.3 \kms). Because \citet{jc90} provide asymmetric drift corrections, 
their RC will be used for the mass models in Sec.~\ref{sec:mass}.

\subsubsection{KAT--7 RC}

\begin{figure}[h!]
\includegraphics[width=\columnwidth]{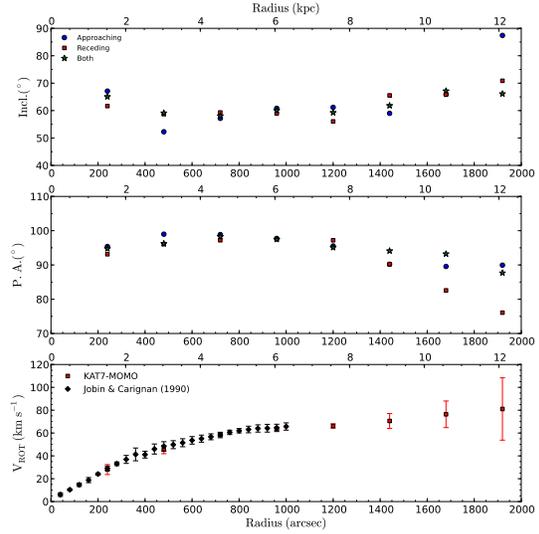}
\caption{Tilted-ring model using the KAT--7 data derived from a moment analysis (intensity weighted)
for both sides and independently 
for the approaching and the receding side and comparison with the RC of  \citet{jc90}.}
\label{fig:kat7_momnt}
\end{figure}

Fig.~\ref{fig:kat7_momnt} shows the tilted-ting model for the intensity-weighted moment analysis of
the KAT-7 na-weighted data of NGC 3109. We find $V_{sys}$ = 405 $\pm$2 \kms , $PA$ $\sim  96^{\rm o}$ $\pm$4
and $i$ $\sim 61^{\rm o}$ $\pm$8. The rotation center is found to be $\sim$ 0.5\arcmin\ North from the optical center
at 10$^{\rm h}$ 03$^{\rm m}$ 06.9$^{\rm s}$ --26$^{\rm o}$ 08\arcmin 58\arcsec. 
This offset from the optical center is not significant and could be due to the larger synthesized beam of KAT--7.  
It can be seen that the agreement between the approaching
and receding sides is much better than for the VLA--ANGST data. The increase in sensitivity allows us to
extend the RC out to $\sim$ 32\arcmin\ ($\sim$12 kpc). Despite the low spatial resolution, no real sign 
of beam smearing is seen when comparing this RC to the VLA data. This may not be surprising in view of the solid-body nature of the RC.

\begin{figure}[h!]
\includegraphics[width=\columnwidth]{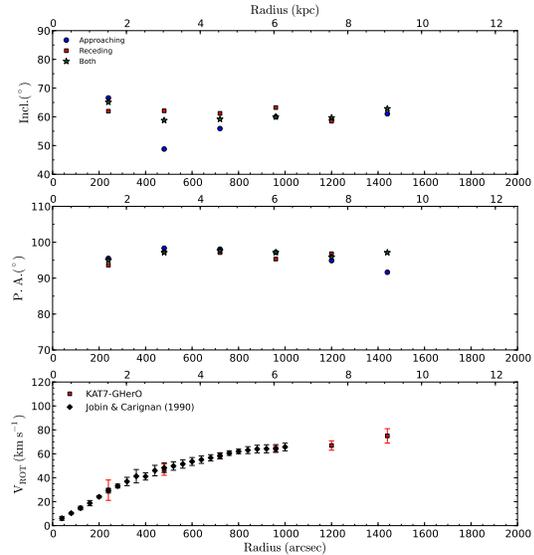}
\caption{Tilted-ring model using the KAT--7 data derived from a Gauss-Hermite polynomials analysis (profile fitting)
for both sides and independently 
for the approaching and the receding side and comparison with the RC of  \citet{jc90}.}
\label{fig:kat7_gh} 
\end{figure}

Fig.~\ref{fig:kat7_gh} shows the tilted-ting model for the Gauss-Hermite polynomials profile fitting analysis of
the KAT-7 na-weighted data. The kinematical parameters found are very similar to those
of the moment analysis with $V_{sys}$ = 406 \kms , $PA$ $\sim 97^{\rm o}$
and $i$ $\sim 61^{\rm o}$ and the same rotation center.
The Gauss-Hermite polynomials are fitted to the spectra in each pixel, where the peak of the fitted profile rises above 5-sigma. As such, profiles from very faint emission at the edge of the galaxy are not strong enough to ensure a good fit. Gauss-Hermite fits will therefore fail at large radii where the average signal-to-noise is lower than the cutoff. Lowering the cutoff produces too many bad velocity points.
Therefore, since the RC using this technique is only defined out to 24\arcmin,
the moment analysis RC is preferred for the mass models in Sec.\ref{sec:mass}.

\subsubsection{Asymmetric drift corrections}

In the case of a galaxy like NGC 3109, where the velocity dispersion represents a substantial fraction of 
the rotational velocity ($\sigma/V_{max} \ge 15-20\%$) and thus provides part of the gravitational support,
a correction for asymmetric drift must be applied. Following the procedure used by \citet{ccf00},
the corrected circular velocity is given by

\begin{equation}
V^2_c = V^2_o - 2\sigma {\delta \sigma \over \delta{\rm ln}\ R} - \sigma^2 {\delta {\rm ln}\ \Sigma \over \delta{\rm ln}\ R}
\end{equation}
where $V_c$ is the corrected velocity, $V_o$ is the observed one, $\sigma$ is the velocity dispersion and $\Sigma$ is the gas surface density.
The asymmetric drift corrections are uncertain by about 25\% \citep{ls89}. 

Table~\ref{KAT7RP} shows the radial profiles for the surface
densities and the velocity dispersion that were derived using the task IRING in AIPS. For the mass models, the \hi\ surface densities
will be multiplied by 4/3 to correct for Helium.
The radial profiles will be used to correct for the asymmetric drift.  Table~\ref{KAT7RC} presents the corrected
values: column (1) gives the radius in arcsec, column (2) the observed rotation velocities, column (3) the errors on those velocities, column (4)
the ratio between the velocity dispersion and the circular velocity, column (5) gives the amplitude of the correction and column (6) the corrected velocities used for the mass models. As can be seen, when $\sigma$/V is greater than 20\%, the corrections can be of the order of a few \kms\, 
which is nearly 10\% for a slow rotator such as NGC 3109. Usually, when $\sigma$/V is less than 20\%, the corrections are of 1 \kms\ or less and usually neglected because they are smaller than the errors.

\begin{table}[h!]
\centering
\caption{Radial variation of the surface densities $\Sigma$ and of the velocity dispersion $\sigma$ for the KAT--7 
data of NGC 3109 from the moment analysis.}
\label{KAT7RP}
\begin{tabular}{ccccc}
\hline\hline
Radius    & $\Sigma$    & $\Delta{\Sigma}$     & $\sigma$ & $\Delta{\sigma}$  \\
(arcsec)   & \msol\ pc$^{-2}$	&   \msol\ pc$^{-2}$            & (\kms)    & (\kms)  \\
\hline
160     &  6.03 & 0.38 & 15.31 & 0.06 \\
374     &  3.44 & 0.31 & 14.32 & 0.15 \\
608     &  2.00 & 0.21 & 12.35 & 0.24 \\
839     &  1.63 & 0.18 & 10.82 & 0.31 \\
1080   &  1.52 & 0.17 & 10.34 & 0.32 \\
1324   &  1.42 & 0.15 &   9.83 & 0.28 \\
1568   &  1.29 & 0.14 &   9.23 & 0.26\\
1807   &  1.02 & 0.11 &   8.56 & 0.27\\
2050   &  0.80 & 0.08 &   8.32 & 0.26\\
2287   &  0.56 & 0.06 &    8.12& 0.23\\
2530   &  0.35 & 0.03 &    7.91& 0.20\\
2765   & 0.22  & 0.02 &    7.27& 0.27\\
\hline
\end{tabular}
\end{table}

\begin{table}[h!]
\centering
\caption{KAT--7 rotation curve of NGC 3109 from the moment analysis, corrected for asymmetric drift.}
\label{KAT7RC}
\begin{tabular}{cccccc}
\hline\hline
Radius    & V$_{\rm 0}$    & $\Delta{\rm V}$     &$\sigma$/V & V$_{\rm corr}$& V$_{\rm C}$ \\
(arcsec)   &	(\kms)	& (\kms)                     & \% & (\kms)   & (\kms)   \\
\hline
240     &  28.0 & 4.4 & 53 & 3.0 &  31.0 \\
480     &  45.5 & 3.6 & 29 & 3.3 &  48.8 \\
720     &  58.5 & 1.5 & 20 & 1.7 &  60.2 \\
960     &  63.9 & 1.2 & 17 & 1.0 &  64.9 \\
1200   &  66.2 & 1.9 & 15 & 0.6 &  66.8 \\
1440   &  70.6 & 6.5 & 14 & 0.8 & 71.4 \\
1680   &  76.4 & 11.6 & 12 & 1.4 &  77.8 \\
1920   &  81.1 & 27.3 & 10 & 0.6 & 81.7\\
\hline
\end{tabular}
\end{table}

\section{Mass models analysis}
\label{sec:mass}

Low mass Surface Density (LSD) galaxies are galaxies whose mass profiles are 
dominated by dark matter (DM) at all galactocentric radii. LSD properties were first
identified in dwarf Irregular (dIrr) galaxies \citep[such as the prototype dIrr DDO 154:][]
{car88,car98} and later in late--type dwarf spirals \citep[see e.g. NGC 5585:][]{ccs91}. 
Despite the uncertainties on the exact M/L ratio of the luminous disk, LSD galaxies, such as 
NGC 3109, are clearly DM dominated at all radii.
For that reason, they can be used to constrain important properties of dark matter
 haloes, such as the characteristic scale density and radius, concentration, virial 
 mass and the exact shape of the mass density profile. Ultimately, measuring the dark matter 
 distribution of these galaxies is necessary if one wants to test the results obtained 
 by numerical simulations of galaxy evolution in the framework of the Cold Dark Matter (CDM) paradigm
 \citep{nfw97} or test alternative gravity theories such as MOND \citep{mil83}.

The study of their mass distribution has generated in the last 15 years the
so-called cusp-core controversy: are rotation curves of LSD
galaxies better reproduced by a cuspy halo as seen in the $\Lambda$CDM numerical 
simulations or by a halo with a nearly constant central density core as seen in most
high spatial resolution observations \citep[e.g.][]{sbo01,db01a,mar02}. A good review
of this debate can be found in \citet{db10}. 

Nowadays, galaxies are expected to form inside cuspy Cold Dark Matter halos.
High resolution velocity fields have provided important observational constraints on the dark 
matter distribution in LSD galaxies. These two-dimensional data show clearly that 
dark matter-dominated galaxies tend to be more consistent with cored than cuspy halos, 
at odds with the theoretical expectations. So, a lot of efforts in the last few years has gone
into identifying the physical processes that could have turned initially cuspy DM halos into
cored ones \citep{gov12,pon12,kuz11,kuz10,gov10}. For example, one recently suggested solution to this problem is to 
enforce strong supernovae outflows that move large amounts of low-angular-momentum gas 
from the central parts and that pull on the central dark matter concentration to create a 
core \citep{fmc13}.

\subsection{Dark Matter models}
\label{sec:dm}

\subsubsection{The pseudo-isothermal DM model (ISO)}
\label{sec:iso}

The pseudo-isothermal DM halo is an observationally motivated model with a constant central density core. 
The density profile is given by:
\begin{equation}
\rho_{ISO}(R) =\frac{ \rho_{0}}{1 + (\frac{R}{R_{c}})^{2}}
\end{equation}
where $\rho_{0}$ is the central density and $R_{c}$   the core radius
\footnote{Note that it is different from the definition of $R_{c}$ for the true isothermal sphere \citep[see][]{kin66,cf85}.}.
The corresponding rotation velocities are given by:

\begin{equation}
V_{ISO}(R) = \sqrt{4\pi G \rho_{0}R_{C}^{2}[1 - \frac{R}{R_{C}}{\rm arctan}(\frac{R}{R_{C}})]}
\end{equation}

We can describe the steepness of the inner slope of the mass density profile with a power law $\rho \sim r^\alpha$.
In the case of the ISO halo, where the inner density is an almost constant density core, $\alpha = 0$.

\subsubsection{The Navarro, Frenk and White DM model (NFW)}
\label{secnfw}

The NFW profile, also known as the "universal density profile" \citep{nfw97} is the commonly adopted dark matter halo 
profile  in the context of the $\Lambda$CDM cosmology. It was derived from N-body simulations. The density profile is given by:
 \begin{equation}
\rho_{NFW}(R) = \frac{\rho_{i}}{R/R_{S}(1 + R/R_{S})^{2}}
\end{equation}
where $R_{S}$ is the characteristic radius of the halo and $\rho_{i}$ is related to the
density of the universe at the time of collapse of the dark matter halo.The corresponding rotation velocities are given by:
\begin{equation}
V_{NFW}(R) = V_{200}\sqrt{ \frac{ln(1 + cx) - cx/(1 + cx)}{x[ln(1 + c) - c/(1 + c)]}}
\end{equation} 
where $x = R/R_{200}$. It is characterized by a concentration parameter $c = R_{200}/R_{S}$ and a velocity $V_{200}$. 
The radius $R_{200}$ is the radius where the density contrast with respect to the critical density of the universe exceeds 200, 
roughly the virial radius \citep{nfw96}. The characteristic velocity $V_{200}$ is the velocity at that radius. The NFW mass 
density profile is cuspy in the inner parts and can be represented by $\rho \sim r^\alpha$, where $\alpha = -1$.

\subsection{ISO \& NFW Models for NGC 3109}
\label{sec:dm3109}

Because of the way the different velocity points are weighted in the mass model fitting
algorithm, we will not combine the high spatial resolution VLA data \citep{jc90} with the
low spatial resolution but high sensitivity KAT--7 data. Instead, we will run a set of models for
each data set. The radial surface density profile of each data set will still be used, keeping in
mind that the VLA data underestimate the \hi\ content, which will not be the case 
for the KAT--7 data.
The I band luminosity profile of \citet{jc90} is preferred to IR (e.g. 2.6$\mu$m)
profiles because it extends to much larger radii. 

\begin{figure}[h!]
\includegraphics[width=\columnwidth]{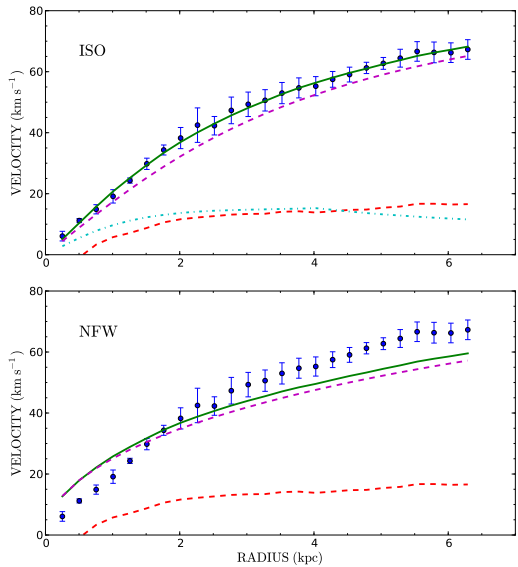}
\caption{DM ISO (top) \& NFW (bottom) mass models for the observed VLA RC (blue points with errors) of NGC 3109 from \citet{jc90}.
The red dash curve is for the \hi\ disk, the dash--dot light blue curve is for the stellar disk,
the purple dash curve is for the DM halo and the continuous green curve is the quadratic sum of the components.
There is no disk component for the NFW model because the best--fit model yields M/L = 0}
\label{fig:dmjc}
\end{figure}

\begin{figure}[h!]
\includegraphics[width=\columnwidth]{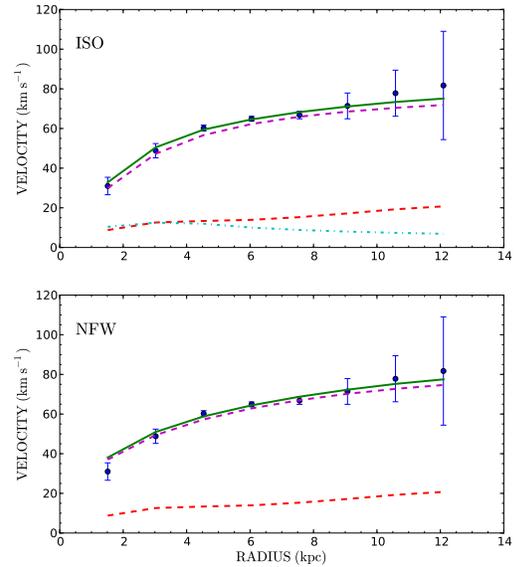}
\caption{DM ISO (top) \& NFW (bottom) mass models for the observed KAT--7 RC (blue points with errors) of NGC 3109.
The symbols are the same as in Fig.~\ref{fig:dmjc}}
\label{fig:dmkat7}
\end{figure}

NGC 3109 presents an interesting test for the DM models. Because it has no bulge 
and a very shallow velocity gradient, it is an ideal system to address the cusp--core controversy.
The DM models are shown in Fig.~\ref{fig:dmjc} for the \citet{jc90} data and in
Fig.~\ref{fig:dmkat7} for the KAT--7 data. The results are summarized in Table~\ref{dmmod}. It can be seen that the ISO
models fit almost perfectly the observed RC with a reduced $\chi^2$ of only 0.24 for the VLA data and 0.31 for the KAT--7 data.
On the other hand, the NFW model has much less success with a reduced $\chi^2$ = 12.9 for the VLA data 
and 0.86 for the KAT--7 data. However, because of the low spatial resolution, the KAT--7 data do not
probe the very inner parts where the discrepancy with the observations is expected to be the greatest.

\begin{table}[h!]
\centering
\caption{Results for the DM mass models of NGC 3109.}
\label{dmmod}
\begin{tabular}{cccc}
\hline\hline
Telescope  & Model & Parameter &  Result \\          
\hline
VLA  & ISO     & $(M/L)_I$ & 0.76 \\
         &              & $R_c$       & 3.2 kpc \\ 
         &              & $\rho_0$   & 0.018 \msol\ pc$^{-3}$ \\
         &              & $\chi^2$    & 0.24 \\
\cline{2-4} 
        & NFW     & $(M/L)_I$ & 0.00 \\
         &              & c                  & 2.1 \\
         &              & $R_{200}$ & 77.4 kpc \\
         &              & $\chi^2$    & 12.9 \\
\hline
KAT--7& ISO & $(M/L)_I$ & 0.55 \\
         &              & $R_c$       & 2.1 kpc \\ 
         &              & $\rho_0$   & 0.028 \msol\ pc$^{-3}$ \\
         &              & $\chi^2$    & 0.31 \\
\cline{2-4}          
         & NFW    & $(M/L)_I$ & 0.00 \\
         &              & c                  & 3.9 \\
         &              & R$_{200}$ & 58.1 kpc \\
         &              & $\chi^2$    & 0.86 \\
\hline
\end{tabular}
\end{table}

The NFW model completely fails in the inner 1 kpc, overestimating the first velocity point by a factor of two for the VLA data. However,
the main problem with the NFW models is that the best fits to both the VLA and the KAT--7 data
suggest a M/L of 0 for the stellar disk, which is unphysical. This is understandable since any stellar disk component would 
just increase the discrepancy in the inner parts.
On the other hand, the best fit ISO model has (M/L)$_I$ = 0.55--0.76 
which is quite compatible with the value predicted by stellar population models of 0.67 $\pm$0.04 \citep{bdj01} for the $I$ band. 
Without any doubt, in the DM halo paradigm, those results confirm that NGC 3109 has a cored and not a cuspy halo.

\subsection{MOND models}
\label{sec:mond}

MOND was proposed by \citet{mil83} as an alternative to
dark matter. Milgrom postulated that at small accelerations the usual Newtonian
dynamics break down and the law of gravity needs to be modified. 
MOND has been claimed to be able to explain the
mass discrepancies in galaxies without dark matter \citep[e.g.][]{bbs91,san96,bot02}.
Therefore, in the MOND formalism, only the contributions of the gas component and of 
the stellar component are required to explain the observed rotation curves. 

 The transition between the Newtonian and the MONDian regime is 
characterized by an acceleration threshold value called a$_{0}$ below which MOND should be used.
So, in the MOND framework, the gravitational acceleration of a test particle is given by :
\begin{equation}
 \mu(x = g/a_{0}) g = g_{N}
\end{equation}
where $g$ is the gravitational acceleration, a$_{0}$ is a new universal constant which
should be the same for all galaxies, $\mu$(x) is the MOND interpolating function and $g_{N}$ the Newtonian acceleration. 

The standard and simple interpolating functions are mostly used in the literature.
 The standard $\mu$-function is the original form of the interpolating function proposed by \cite{mil83}. However,  
 \cite{zf06} found that a simplified form of the interpolating function not only provides also good 
 fits to the observed rotation curves but the derived mass-to-light ratios 
 are more compatible with those obtained from stellar populations synthesis models.

\subsubsection{MOND models using the "standard" interpolation function}
\label{sec:standard}

The standard interpolating function is given as
\begin{equation}
\mu(x) = \frac{x}{\sqrt{1 + x^{2}}}
\end{equation}
For $x \ll 1$ the system is in deep MOND regime with $g = (g_{N}$a$_{0}$)$^{1/2}$ and for $ x \gg$ 1 the gravity is Newtonian.\\

The MOND rotation curve becomes:

 \begin{equation}
V_{rot}^{2} = \frac{V_{sum}^{2} }{\sqrt{2}}\sqrt{1 + \sqrt{1 + (2ra_{0}/V_{sum}^{2})^{2}}}
\end{equation}

where 
\begin{equation}
V_{sum}^{2} = V_{b}^{2}+ V_{d}^{2}  + V_{g}^{2}
\end{equation}
$V_{b}$, $V_{d}$, $V_{g}$ are the contributions from the  bulge, the disk and the gas to the rotation curve. In the case of the
Magellanic-type spiral NGC 3109, there is no bulge to consider.

\subsubsection{MOND models using the "simple" interpolation function}
\label{sec:simple}

The simple interpolating function is given by
\begin{equation}
\mu(x) = \frac{x}{1 + x}
\end{equation}

 Using the same procedure as in previous section we can easily obtain the corresponding rotation velocities:
 \begin{equation}
V_{rot}^{2} = \sqrt{V_{b}^{2}+ V_{d}^{2}  + V_{g}^{2}}*\sqrt{ a_{0}*r + V_{b}^{2}+ V_{d}^{2}  + V_{g}^{2}}
\end{equation}

\subsection{MOND Models for NGC 3109}
\label{sec:mond3109}

NGC 3109 presents an interesting test for MOND. As we have seen, it is close enough to have a very accurate
distance determination using Cepheids and the largest portion of the luminous mass is in the form of gas and not
stars, partly freeing us from the uncertainties due to the M/L value used for the disk. The internal accelerations are very low,
therefore the galaxy is completely within the MOND regime \citep{la89}. 

\begin{figure}[h!]
\includegraphics[width=\columnwidth]{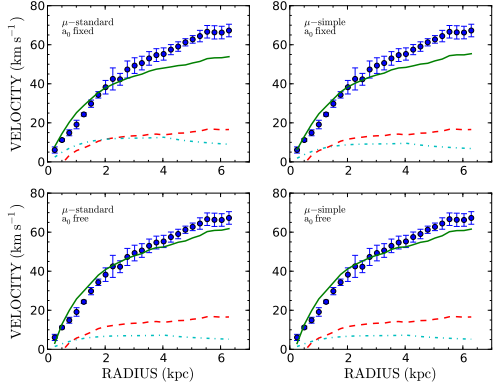}
\caption{MOND mass models with $a_0$ fixed (top) and $a_0$ free (bottom) and for the standard (left) and the simple (right) interpolation functions for NGC 3109 using the RC of \citet{jc90}. The red dash curve is for the \hi\ disk, the dash--dot light blue curve is for the stellar disk,
 and the continuous green curve is the MOND model.}
\label{fig:mondjc}
\end{figure}

\begin{figure}[h!]
\includegraphics[width=\columnwidth]{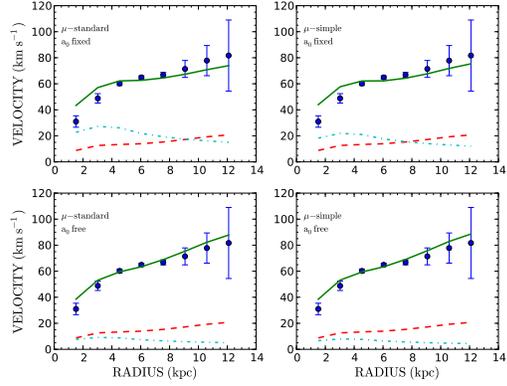}
\caption{MOND mass models with $a_0$ fixed (top) and $a_0$ free (bottom) and for the standard (left) and the simple (right) interpolation functions for NGC 3109 using the RC from KAT--7 data. The symbols are the same as in Fig.~\ref{fig:mondjc}}
\label{fig:mondkat7}
\end{figure}

The MOND mass models for the VLA data are presented in Fig.~\ref{fig:mondjc} \& Table~\ref{mondmodVLA}
and in Fig.~\ref{fig:mondkat7} \& Table~\ref{mondmodKAT7} for the KAT--7 data,
for both the {\it standard} value of $a_0$, namely
1.21 x 10$^{-8}$ cm sec$^{-2}$ \citep{bbs91} and with $a_0$ left as a free parameter. The {\it standard}
and the {\it simple} interpolation functions are also illustrated. 

\begin{table}[h!]
\centering
\caption{Results for the MOND models of NGC 3109 for the VLA data.}
\label{mondmodVLA}
\begin{tabular}{cccc}
\hline\hline
 a$_0$ & $\mu$ & Parameter & Result \\
  10$^{-8}$ cm s$^{-2}$ & & & \\          
\hline
fixed    & standard & (M/L)$_{\rm I}$ & 0.45 \\
                       &                  & a$_0$                 & 1.21 \\
                        &                  &$\chi^2$              & 12.01 \\
\cline{2-4}
               & simple     & (M/L)$_{\rm I}$ & 0.26 \\
                       &                  & a$_0$                 & 1.21 \\
                        &                  &$\chi^2$              & 9.25 \\
\cline{2-4}       
          free    & standard & (M/L)$_{\rm I}$ & 0.15 \\
                        &                  & a$_0$                 & 2.48 \\
                        &                  &$\chi^2$              & 5.50 \\
\cline{2-4}
             & simple     & (M/L)$_{\rm I}$ & 0.15 \\
                        &                  & a$_0$                 & 2.07 \\
                        &                  &$\chi^2$              & 5.57 \\
\hline
\end{tabular}
\end{table}

\begin{table}[h!]
\centering
\caption{Results for the MOND models of NGC 3109 for the KAT--7 data.}
\label{mondmodKAT7}
\begin{tabular}{cccc}
\hline\hline
 a$_0$ & $\mu$ & Parameter & Result \\
  10$^{-8}$ cm s$^{-2}$ & & & \\          
\hline

 fixed    & standard & (M/L)$_{\rm I}$ & 2.61 \\
                           &                  & a$_0$                 & 1.21 \\
                           &                  &$\chi^2$              & 2.96 \\
\cline{2-4}
                    & simple     & (M/L)$_{\rm I}$ & 1.69 \\
                             &                  & a$_0$                 & 1.21 \\
                             &                  &$\chi^2$              & 3.38 \\
\cline{2-4}
                free    & standard & (M/L)$_{\rm I}$ & 0.29 \\
                             &                  & a$_0$                 & 3.48 \\
                             &                  &$\chi^2$              & 1.30 \\
\cline{2-4}
                    & simple     & (M/L)$_{\rm I}$ & 0.22 \\
                             &                  & a$_0$                 & 3.25 \\
                             &                  &$\chi^2$              & 1.37 \\
\hline
\end{tabular}
\end{table}

In the case of the VLA data, the fits are poor for all the cases:  they overestimate the velocities in the inner parts and they underestimate them in the 
outer parts. The reduced ${\chi}^2$ are much larger than for the DM fits, varying from 5.5 to 12. However, since most of the
luminous mass is in gas and since the VLA misses 40\% of the flux, we should not expect good fits.

The MOND models have more significance for the KAT--7 data than for the VLA data, since we probe the gas on all scales. In this case we see that
for the standard fixed value of a$_0$ = 1.21 x 10$^{-8}$ cm sec$^{-2}$, the results of the models are poor with reduced $\chi^2$ of 3.0 to 3.4 and
(M/L)$_B$ values 3 to 4 times larger than the value predicted by stellar population models. However, we see that if a$_0$ is free to vary, we get more
reasonable fits with reduced $\chi^2$ down to 1.30 and 1.37, still 4--5 times larger than for the DM ISO models. The main problems with those
models are the very small (M/L)$_I$ values, 2 to 3 times smaller than the expected values and the large values of the constant $a_0$ which is
greater than a factor of 2 compared to the standard value.

\section{Discussion}
\label{sec:discussion}

We will now examine different questions that were raised by this study. The first one
is the difference found between the kinematic inclination and the photometric 
values. This is an important point to understand since $i$ scales the RC and is responsible for the velocity
gradient of the rising part of the RC, which really constrains the mass models and set the M/L ratio of the luminous disk.
Both tilted-ring models using the intensity-weighted and the profile fitting techniques find the same inclination of $i$ = 61$^{\rm o}$.
It is important to understand how this parameter is determined. Assuming an axisymmetric disk, the tilted--ring model
tries to find the value that minimizes the dispersion of the velocities in the rings.

On the other hand, looking at Table~\ref{optorpar}, it can be seen that the inclination varies from 78.5$^{\rm o}$
in B to 73.5$^{\rm o}$ in I and 69.5$^{\rm o}$ in the 3.6$\mu$m. What is going on is quite clear. Each photometric band samples different stellar
populations. The B--band samples the young blue Pop I stars which are confined to the thin disk, the I--band samples
a mixture of young and old Pop I stars which are in a thicker disk while the 3.6$\mu$m is completely dominated by
an old disk population in an even thicker disk. This explains that progression of the inclination from $\sim$80$^{\rm o}$
to $\sim$70$^{\rm o}$. 

\begin{table}[h!]
\centering
\caption{Optical orientation parameters of NGC 3109.}
\label{optorpar}
\begin{tabular}{lcl}
\hline\hline
Parameter  & & Ref \\          
\hline
Orientation parameters, $B$ &&\\
Axis ratio, b/a & 0.20 $\pm$ 0.02 & (1) \\
Inclination\tablenotemark{*}                 &  78.5$\rm ^o$  $\pm\ 2.0^{\rm o}$                      & (1)   \\
\pa                                &  91.5$\rm ^o$   $\pm\ 1.0^{\rm o}$              & (1)   \\
Orientation parameters, $I$ &&\\
Axis ratio, b/a & 0.28 $\pm$ 0.02 & (2) \\
Inclination\tablenotemark{*}                   &  73.5$\rm ^o$   $\pm\ 1.5^{\rm o}$                     & (2)   \\
\pa                                &  93.0$\rm ^o$   $\pm\ 2.0^{\rm o}$                     & (2)   \\

Orientation parameters, 3.6 $\mu$m &&\\
Axis ratio, b/a& 0.35 $\pm$ 0.02								&  (3)   \\
Inclination\tablenotemark{*}                   &  69.5$\rm ^o$  $\pm\ 1.5^{\rm o}$                      & (3)   \\
\pa                                &  95.0$\rm ^o$   $\pm\ 2.0^{\rm o}$                    & (3)   \\
\hline
\tablenotetext{*}{no correction for intrinsic flattening.}
\tablerefs{(1)  \citet{car85}; \\ (2) \citet{jc90}; (3) this study.}
\end{tabular}
\end{table}

It is instructive to look at the ellipse fits to the \hi\ isophotes. As can be seen in Table~\ref{HIISO}, NGC 3109 is really
composed of 2 disks: an inner one, corresponding to the optical disk, 
with $i \sim$76$^{\rm o}$, which is the mean of the B and I inclinations and an outer
disk with $i \sim$63$^{\rm o}$ which, within the errors, agree with the kinematical inclination.
The double disk is quite apparent when looking at the \hi\ isophotes in Fig.~\ref{fig:kat7_squash_B}.
Therefore, since the photometric inclination is most sensitive to the emission of the stellar populations being traced, it is 
quite clear that the kinematic inclination should be preferred. The photometric values should only be used as starting
values in the tilted--ring modeling.

\begin{table}[h!]
\centering
\caption{Fits to the \hi\ isophotes.}
\label{HIISO}
\begin{tabular}{ccccc}
\hline\hline
Isophotes    & q (axis ratio)    & $\Delta$q     & i (inclination) & $\Delta$i  \\
\hline
1.00 	&	0.506 &	0.049 & 59.6 & 3.3 \\
3.00 	&	0.403 &	0.022& 66.2 & 1.4 \\
5.00 	&	0.256 &	0.029 & 75.2 & 1.8 \\
7.00 	&	0.246 &	0.018 & 75.8 & 1.1 \\
9.00 	&	0.257 &	0.011 & 75.1 & 0.7 \\
11.00 &	0.255 &	0.007 & 75.2 & 0.3 \\
13.00 &	0.247 &	0.005& 75.7 & 0.3 \\
15.00 &	0.247 &	0.004& 75.7 & 0.2 \\
17.00 &	0.255 &	0.004& 75.2 & 0.2 \\
19.00 &	0.278 &	0.005& 73.9 & 0.3 \\
21.00 &	0.326 &	0.006 & 71.0 & 0.4 \\
23.00 &	0.448 &	0.011& 63.4 & 0.7 \\
25.00 &	0.460 &	0.016& 62.6 & 1.0 \\
27.00 &	0.460 &	0.018& 62.6 & 1.2 \\
29.00 &	0.446 &	0.018& 63.5 & 1.2 \\
31.00 &	0.446 &	0.010 & 63.5 & 0.6 \\
\hline&
\end{tabular}
\end{table}

Another point worth discussing is beam-smearing. With such a large synthesized beam $\sim$4\arcmin ,
one would have expected beam smearing to be quite important. However when looking at Figures~\ref{fig:kat7_momnt}
and \ref{fig:kat7_gh}, this does not seem to be the case. 
As explained in \citet{car85}, beam smearing is the result of the convolution of the Gaussian beam  with the \hi\
distribution and the velocity gradient across the beam. If the \hi\ distribution is steep and/or the
velocity gradient is large, the net effect will be to underestimate the observed velocity or conversely
to overestimate the effective radius (the product of the convolution) that is observed. In the case of NGC 3109,
the \hi\ distribution is fairly flat across the beam and the velocity gradient of that galaxy is very small.
Those two properties render the beam smearing negligible.
Should we be observing another galaxy with a step velocity gradient and/or a steep radial \hi\  distribution, 
beam smearing would be important.

Let us now turn to the mass models. First, the results of the DM ISO models for both RCs confirm the 
previous results that it provides an almost perfect fit to the observed kinematics with the difference that the
KAT--7 model has a more massive \hi\ disk. As expected, this translates in a smaller stellar disk M/L value for the KAT--7 data.
In both cases, the mass--to--light ratio found for the disk is compatible with population 
synthesis models. The NFW models again fail to reproduce the kinematics.
In the inner kpc, the velocities are overestimated by a large factor, despite an unphysical M/L ratio of 0 for the stellar disk. 
In the DM halo paradigm, clearly NGC 3109 has a cored and not a cuspy halo, at least at the present epoch.

What about the MOND models ? Twenty-five years ago, 
 \citet{san86} pointed out that the MOND mass of NGC 3109 predicted using Milgrom's suggested value for a$_{\rm 0}$
was 5 x 10$^8$ \msol , thus larger than the \hi\ mass using the larger distance estimates known at the time
between 1.7 to 2.6 Mpc \citep{car85}. This is not the case anymore with the smaller well determined distance of 1.3 Mpc,
but still our determined \hi\ mass with KAT--7 is less than 10\% smaller than this MOND mass, which implies 
unphysically small M/L values for the stars in the case where the fits have been improved letting the constant $a_0$ free to vary. 

When \citet{bbs91} produced a MOND model
using the \citet{jc90} VLA data, they argued that they could not get a good fit because the VLA data was missing a
substantial part of the flux and that the \hi\ mass had to be multiplied by 1.67 to get a reasonable fit. However, 
since the KAT--7 data retrieve all the NGC 3109 flux, this cannot be used as an argument with the present data
which convincingly show that MOND cannot reproduce the observed kinematics of NGC 3109
with physically acceptable parameter values. Unless
some other explanation can be found, the KAT--7 data surely challenge the MOND theory.

What about the possible interaction between the Magellanic-type spiral NGC 3109 and the dIrr/dSph Antlia,
suggested by the \hi\ isophotes being slightly elongated ? 
The high surface brightness sensitivity of the KAT-7 observations allow us to trace the lopsidedness 
of NGC 3109 to larger radii, however there is no further obvious evidence of an interaction 
with Antlia, leaving the question open for further investigation.
NGC 3109 has a systemic velocity of 
404 \kms\ while Antlia has 360 \kms. \citet{ap97} calculated the physical separation between NGC 3109 and Antlia 
to be between 29 and 180 kpc, with a maximum separation of 37 kpc for the pair to be bound. 

The situation for the Local Group dwarfs can guide us. \citet{gp09} showed clearly that the majority of dwarf
galaxies within 270 kpc of the Milky Way or Andromeda are undetected in \hi\ ($< 10^4$ \msol\ for the Milky Way
dwarfs), while those further than 270 kpc are predominantly detected with masses $10^5$ to $10^8$ \msol\
(Antlia has an \hi\ mass of $\sim$ 10$^6$ \msol)
meaning the nearby ones must have been stripped of their gas. While the halo of NGC 3109 is not
as large as the Milky Way halo, if a close encounter had happened in the past, it is quite likely that Antlia would also have been stripped of its gas. 

Nevertheless, some kind of interaction really seems to have taken place, as shown by the \hi\ isophotes of the two galaxies 
pointing toward each other. However, in view of the large difference in masses between the two systems and the absence 
of any external traces of such an interaction, it is believed that the internal kinematics of NGC 3109 cannot have been 
severely altered.

Anyway, if such an interaction did take place, it would be more the rule than the exception. Massive galaxies 
usually have a significant population of gas-rich dwarf companions and interaction with these will show kinematic and
morphological signatures in the extended HI disks \citep{mih12} such as warps, plumes, tidal tails, high-velocity
clouds \citep[e.g.][]{hib01,san08} or even stellar streams \citep{lew13}. Galaxies are said to be embedded in the cosmic web, 
seen in the $\Lambda$CDM numerical simulations \citep{sfw06}. Such signatures of interaction have been studied close 
to massive galaxies such as the interconnecting network in the M81/M82 system \citep{yhl94} or the tidal tail in the Leo triplet 
\citep{hgr79}. More recently, \citet{mih12} using the Green Bank Telescope (GBT) found a plume in the outer disk of
M101 with a peak column density of 5 x 10$^{17}$ cm$^{-2}$ and two new \hi\ clouds close to that plume with masses of 
$\sim$10$^7$ \msol. While KAT--7 would not have detected such low column densities, it is interesting to look at our detection 
cloud mass limit.

To calculate a characteristic \hi\ mass sensitivity, we assume that low--mass clouds would be unresolved in our beam and use the relation \citep{mih12}:

\begin{equation}
({{\sigma_M} \over{M\odot}}) = 2.36 \times 10^5 ({D^2 \over {\rm Mpc}}) ({{\sigma_s} \over {\rm Jy}}) ({{\Delta V} \over {\rm km\ s^{-1}}})
\end{equation}
where $D$ is the distance, $\sigma_s$ is the rms noise in one channel and $\Delta V$ is the channel width. This means that 
our 3$\sigma$ cloud mass detection limit is around 5--6 x 10$^3$ \msol\ at the distance of NGC 3109. We should thus have 
easily detected similar clouds like the ones observed around M101, if they had been present.

\section{Summary and conclusions}
\label{sec:conclusion}

The first \hi\ spectral line observations with the prototype radio telescope KAT--7 have been presented. The
high sensitivity of KAT--7 to large scale, low column density emission comes not only from its compact configuration, 
but also from its very low T$_{sys}$ receivers. With $\sim$25 hours of observations per pointing,
surface densities of 1.0 x 10$^{19}$ atoms cm$^{-2}$ were reached, which could be improved when 
the telescope will be fully
commissioned, since the theoretical noise was not reached with the present dataset.

The main results from this study are:
\begin{itemize}
\item A total \hi\ mass of 4.6 x 10$^8$ \msol\ is measured for NGC 3109, using our adopted distance of 1.3 Mpc. 
This \hi\ mass, which is $\sim$40\% larger than the values calculated using VLA observations, is
surely a better estimate of the total \hi\ mass of NGC 3109 since KAT--7 is sensitive to the large scales for
which the VLA is not.
\item The \hi\ disk extends over a region of 58\arcmin (EW) x 27\arcmin (NS) down 
to a limiting column density of 1.0 x 10$^{19}$ atoms cm$^{-2}$.
\item The \hi\ distribution is lopsided with more \hi\ on the SW approaching side. Because of this, no 
intensity-weighted velocity is derived from the global profile but rather a
mid-point velocity of 404 $\pm$ 2 \kms , believed to be more representative of the systemic velocity of NGC 3109.
Profile widths of $\Delta$V$_{50}$ = $118 \pm 4$ \kms\ and $\Delta$V$_{20}$ = $133 \pm 3$ \kms\ are derived.  
\item VLA-ANGST data were used to derive the \hi\ properties of the dSph/dIrr Antlia, since the spatial resolution
of KAT--7 is too low to study this dwarf system.
A total \hi\ mass of 1.5 x 10$^6$ \msol\ is measured for our adopted distance of 1.31 Mpc. 
An intensity-weighted mean velocity of $360 \pm 2$ \kms\ is derived along with a $\Delta$V$_{50}$ = 
$23 \pm 3$ \kms\ and a $\Delta$V$_{20}$ = $33 \pm 3$ \kms. 
\item A tilted--ring model was obtained for the high velocity resolution VLA--ANGST data of NGC 3109. As on the
\hi\ total intensity map, it can be seen that the warp is more important on the approaching side. The RC found is nearly
identical to the one derived by \citet{jc90} which had a velocity resolution of $\sim$10 \kms. Because the latter study provides 
correction for asymmetric drift, it was preferred for the mass model analysis.
\item Rotation curves from the KAT--7 data of NGC 3109 were derived from two different types of analysis: an 
intensity--weighted moment analysis and a Gauss--Hermite polynomial profile fitting. Both data sets give essentially the same
result with $V_{sys}$ = 405 \kms\ and mean $PA$ = 96$^{\rm o}$ and $i$ = 61$^{\rm o}$, with  very little difference between the
approaching and the receding side. The KAT--7 RCs agree very well with the VLA data in the inner parts 
while allowing to extend the rotation data by a factor of 2 out to 32\arcmin. Since the moment analysis data allow us to derive the 
RC further out, it was used for the mass models.
 \item The observationally motivated  DM ISO model reproduces very well the observed RCs of both
 the \citet{jc90} and the KAT--7 data while a cosmologically motivated 
 NFW model gives a much poorer fit, especially in the very inner parts.
 Because of the high spatial and velocity resolutions data available and the very small errors on those velocities, this
 cannot be attributed to poor data. NGC 3109 definitely has a cored and not a cuspy DM halo.
 \item While it is clear that having the proper gas distribution has reduced the discrepancies between the observed RC and the MOND models,
 the unreasonable (M/L) and large $a_0$ values obtained lead us to conclude that we cannot get acceptable MOND models for NGC 3109.
 The distance being so well determined with very small errors from Cepheids observations 
 and the \hi\ mass so well constrained by the KAT--7 observations, uncertainties on these two values cannot explain why the MOND models fail
 to reproduce the observed kinematics with reasonable parameters.
 \item Besides some elongation of the outer isophotes, already seen in previous observations, 
 no further evidence is found for past 
encounter and/or interaction between the Magellanic-type spiral NGC 3109 and the dSph/dIrr Antlia.
\end{itemize}

Our findings for NGC 3109 are not an isolated case. Recently, \citet{san13} studied a sample of slowly rotating gas--rich galaxies in the 
MOND framework. These are again galaxies in the full low acceleration MOND regime. They found at least five such systems (especially 
NGC 4861 and Holmberg II) that deviate strongly from the MOND predictions, unless their inclinations and distances differ strongly from 
the nominal values. In the case of NGC 3109, those two parameters are much more constrained which makes it harder to reconcile MOND 
with the observed kinematics.

Those observations obtained with KAT-7 have shown that despite its relatively small collecting area
(7 x 12 m antennae), this telescope really has a niche for detecting large scale low emission over
the $\sim$1$^{\rm o}$ FWHM of its antennae. It should be kept in mind that this telescope was built primarily as 
a testbed for MeerKAT and the SKA such that any scientific result that can be obtained is a bonus.
While most of the extragalactic \hi\ sources would be unresolved
by the $\sim$4\arcmin\ synthesized beam, many projects such as this one on NGC 3109 can be
done on nearby very extended objects such as Local Group galaxies or galaxies in nearby groups like Sculptor. 

\acknowledgements

We thank all the team of SKA South Africa for allowing us to get scientific data during the 
commissioning phase of KAT--7 and are grateful to the ANGST team for making their reduced VLA data publicly available. 
CC's work is based upon research supported by the South African Research Chairs Initiative (SARChI) of the Department of Science 
and Technology (DST),  the Square Kilometer Array South Africa (SKA SA) and the National Research Foundation (NRF).
The research of BF, KH, DL \& TR have been supported by SARChI, SKA SA and National Astrophysics and Space Science Programme
(NASSP) bursaries.

\appendix

\section{Appendix: Other galaxies in the field}

\subsection{ESO 499-G037 (UGCA 196)}

\begin{figure}
\includegraphics[scale=0.5]{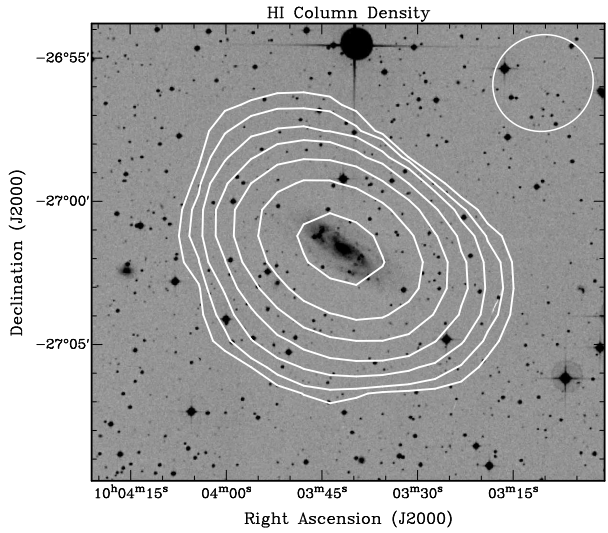}
\caption{KAT--7 data on ESO 499-G037.  The \hi\ emission is superposed on a DSS B image
The contours are 0.1, 0.2, 0.4, 0.8, 1.6, 3.2, 6.4 $\times 10^{20}$ atoms cm$^{-2}$.}
\label{fig:G037_mom0}
\end{figure}

ESO 499-G037 is a SAB(s)d spiral \citep{dev91}. It has very bright \hii\ regions and star formation activity, as seen in
the GALEX image \citep{dep07}. Fig.~\ref{fig:G037_mom0} shows the \hi\ intensity map, superposed on an optical image. 
We see that the \hi\ disk has a diameter of
$\sim$12.5\arcmin , nearly 4 times the optical size \citep{dev91}. Because of the low spatial resolution of the KAT--7 data, 
these observations do not allow to derive a proper RC. 

\begin{figure}
\includegraphics[scale=0.4]{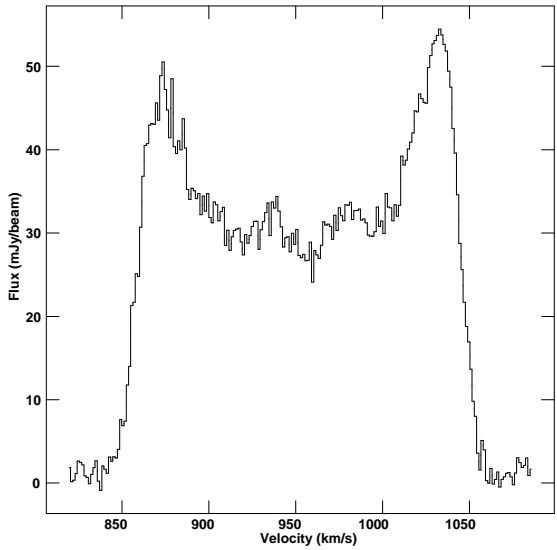}
\caption{Global \hi\ profile for ESO 499--G037.}
\label{fig:G037_gp}
\end{figure}

Fig.~\ref{fig:G037_gp} shows the global profile of ESO 499-G037. From it, a systemic velocity of 953 $\pm$ 3 \kms\
is found, similar to the HIPASS determination of 954 $\pm$5 \kms\ \citep{kor04} and the value of 955 $\pm$ 1 \kms\ 
determined by  \citet{bdb01}. 
Velocity widths of $\Delta$V$_{50}$ = $184 \pm 4$ \kms\ and $\Delta$V$_{20}$ = $200 \pm 4$ \kms\ are derived.
A total flux of 51.7 $\pm$ 5.2 Jy \kms\ is measured, which is $\sim$25\% larger than the HIPASS flux of 40.2 $\pm$ 4.2 Jy \kms\
but consistent with the value of 49 $\pm$ 2 Jy \kms\ from \citet{bdb01}, also
using the Parkes multi--beam system, or the Green Bank value of 48.4 $\pm$ 2.4 Jy \kms\ \citep{spr05}.

\subsection{ESO 499-G038 \& the \hi\ cloud}

\begin{figure}
\includegraphics[scale=0.75]{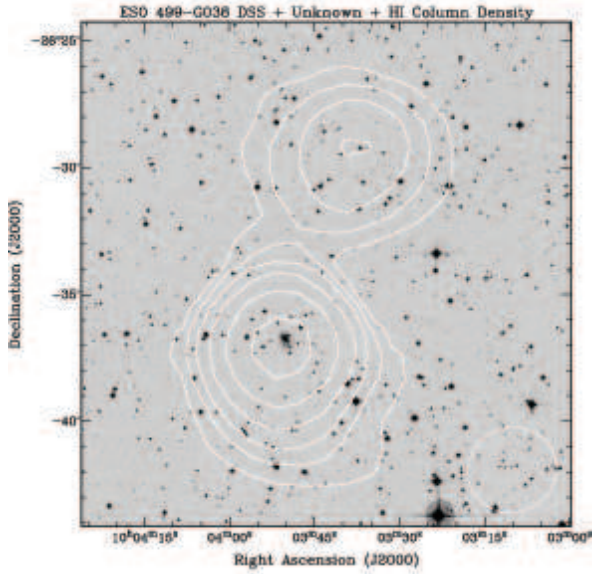}
\caption{KAT--7 data on ESO 499-G038. The \hi\ emission of the galaxy and the cloud to the NW are
superposed on a DSS B image.
The contours are 0.025, 0.1, 0.2, 0.4, 0.8, 1.6 $\times 10^{20}$ atoms cm$^{-2}$.}
\label{fig:G038_mom0}
\end{figure}

\begin{figure}
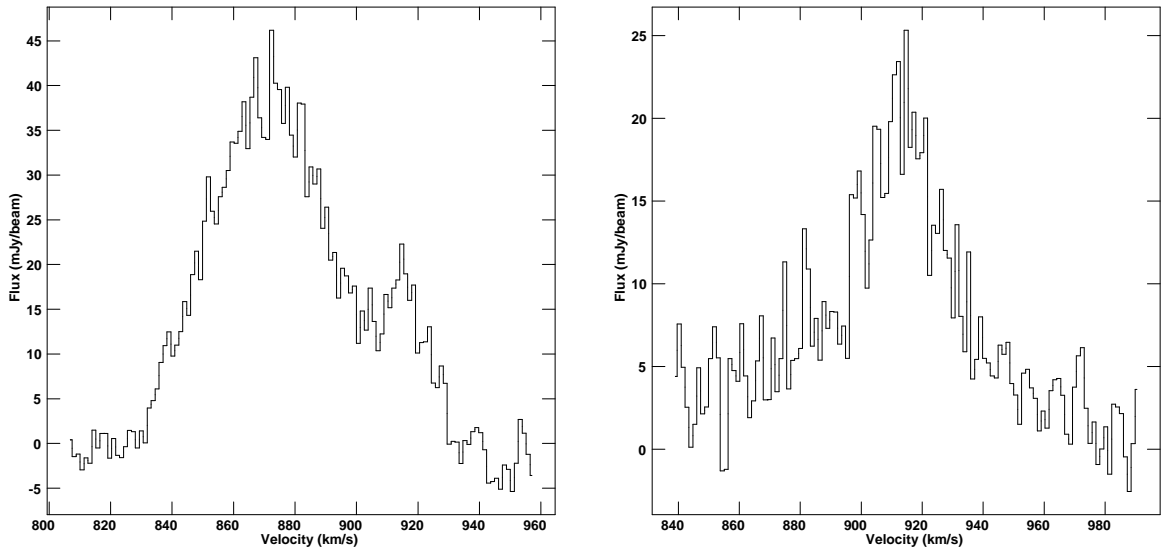

\centering
\begin{tabular}{cc}
\includegraphics[scale=0.4]{fig19a.eps}&
\includegraphics[scale=0.4]{fig19b.eps}\\
\end{tabular}
\caption{Global \hi\ profile for ESO 499--G038 ($\sim$ 870 \kms) and the cloud to the NW ($\sim$ 910 \kms).}
\label{fig:G038_gp}
\end{figure}

ESO 499-G038 is a late--type Sc galaxy with, as can be seen in Fig.~\ref{fig:G038_mom0}, a very extended \hi\ component and
a likely associated \hi\ cloud to the NW, as shown by the faintest contour at 2.5 x 10$^{18}$ cm$^{-2}$.  
While the velocity field shows clear rotation, it is not
possible to say more about the kinematics with the present data. Our spatial resolution is sufficient, however, to nearly
resolve the two components in the global \hi\ profile shown in Fig.~\ref{fig:G038_gp}. 

We measured a systemic velocity of 871 $\pm$ 4 \kms\ for ESO 499-G038 and 912 $\pm$ 5 \kms\ for the \hi\ cloud. These are surely better estimates
then the values of 885 $\pm$ 5 \kms\ \citep{kor04} and 888 $\pm$ 3 \kms\ \citep{bdb01} given for the systemic velocity of ESO 499-G038
from their unresolved multi---beam profiles. Fluxes of 9.4 $\pm$0.9 and 1.5 $\pm$ 0.2 Jy \kms\ are measured for the galaxy and the cloud.
This can be compared to 9.5 $\pm$ 2 \citep{kor04} and 11.4 $\pm$1 Jy \kms\ \citep{bdb01} from the unresolved Parkes' spectra.

\end{document}